\documentclass[manuscript]{emulateapj}

\usepackage{graphics}
\usepackage{natbib}
\usepackage{amsmath}
\usepackage[colorlinks,urlcolor=blue,citecolor=blue,linkcolor=blue]{hyperref}

\shortauthors{Tamayo et al.}

\begin{document}

\title{dynamical stability of imaged planetary systems in formation: \\ application to hl tau}

\author{D. Tamayo\altaffilmark{1,2,3}, A. H. M. J. Triaud\altaffilmark{1,3,4}, K. Menou\altaffilmark{1,4} and H. Rein\altaffilmark{1}}
\altaffiltext{1}{Department of Physical \& Environmental Sciences, University of Toronto at Scarborough, Toronto, Ontario M1C 1A4, Canada}
\altaffiltext{2}{Canadian Institute for Theoretical Astrophysics, 60 St. George St, University of Toronto, Toronto, Ontario M5S 3H8, Canada}
\altaffiltext{3}{Centre for Planetary Sciences Fellow}
\altaffiltext{4}{Department of Astronomy \& Astrophysics, University of Toronto, Toronto, Ontario M5S 3H4, Canada}
\email{d.tamayo@utoronto.ca}

\begin{abstract}
A recent ALMA image revealed several concentric gaps in the protoplanetary disk surrounding the young star HL Tau.  We consider the hypothesis that these gaps are carved by planets, and present a general framework for understanding the dynamical stability of such systems over typical disk lifetimes, providing estimates for the maximum planetary masses.  We collect these easily evaluated constraints into a workflow that can help guide the design and interpretation of new observational campaigns and numerical simulations of gap opening in such systems.  We argue that the locations of resonances should be significantly shifted in massive disks like HL Tau, and that theoretical uncertainties in the exact offset, together with observational errors, imply a large uncertainty in the dynamical state and stability in such disks.  This presents an important barrier to using systems like HL Tau as a proxy for the initial conditions following planet formation.  An important observational avenue to breaking this degeneracy is to search for eccentric gaps, which could implicate resonantly interacting planets.  Unfortunately, massive disks like HL Tau should induce swift pericenter precession that would smear out any such eccentric features of planetary origin.  This motivates pushing toward more typical, less massive disks.  For a nominal non-resonant model of the HL Tau system with five planets, we find a maximum mass for the outer three bodies of approximately 2 Neptune masses.  In a resonant configuration, these planets can reach at least the mass of Saturn.  The inner two planets' masses are unconstrained by dynamical stability arguments.

\end{abstract}


\section{Introduction}
Given its wavelength coverage, sensitivity and resolution, the Atacama Large Millimeter/Submillimeter Array (ALMA) represents a unique observatory for studying not only the extent and masses of protoplanetary disks, but also their structure.  The latter provides the possibility of detecting forming planets as they gravitationally sculpt the surrounding disk.  From October 24-31 2014, as part of its science verification process, the ALMA team pointed 25-30 antennas at the young star HL Tau (J2000 04:31:38.45 +18:13:59.0), measuring the continuum emission in Band 6 (211-275 GHz), with baselines from 15 m to 15 km\footnote{\url{https://public.nrao.edu/news/pressreleases/planet-formation-alma}}.  As of the time of this writing, no scientific publication has appeared analysing these data; however, the ALMA project has publicly released an image of the system (see Fig.\:\ref{hltauimg}) that reveals several concentric dust gaps, perhaps revealing forming planets massive enough to sculpt the nearby disk material.  

This striking discovery during the science verification phase suggests that ALMA may find similar features in several other young systems.  In this paper, we therefore generally investigate the constraints one can extract on potential planetary masses from the requirement that observed systems be dynamically stable.  We then apply these constraints to the HL Tau system using approximate gap separations from the publicly released image.  As argued in Appendix \ref{Amaury}, our analysis should be largely insensitive to an improved determination of the orbital radii.

We begin in Sec.\:\ref{ass} with a list of assumptions that we make in our analysis.  Section \ref{inst} then investigates stability in non-resonant systems and applies our thresholds to numerical simulations of the HL Tau system, finding that resonances may be important in shaping the system's stability.  Section \ref{res} discusses the role of resonances, and provides simple criteria for evaluating whether resonances must be considered for stability.  Section \ref{upper} outlines a procedure for numerically investigating resonant systems, applying it to HL Tau.
Section \ref{proc} provides a summarized workflow for how one might analyze a potential planetary system embedded in a gas disk, making reference to more detailed discussion in the main text.  Our conclusions for the HL Tau system can be found in Sec.\:\ref{hltauconc}.    

\section{Assumptions} \label{ass}
By specializing to protoplanetary disks observed by ALMA, we can restrict the relevant range of scales, simplifying our analysis.  For any pair of putative planets (throughout the paper, subscripts 1 and 2 to the inner and outer body, respectively) we assume
\begin{eqnarray}
\frac{\Delta a}{a_1} \gtrsim 10^{-1}, \label{da} \\
P_{1,2} \gtrsim 10 \: \text{yrs}, \label{P} \\
\frac{M_{1,2}}{M_\star} \lesssim 10^{-2}, \label{M} \\
t \lesssim 1 \text{Myr}
\end{eqnarray}
where $a$, $P$, and $M$ represent the semimajor axis, orbital period, and mass, respectively; $\Delta a$ is the orbital separation, the $\star$ subscript refers to the central star, and $t$ is the age of the system.  Since the maximum resolution of ALMA is $\gtrsim 10$ milliarcseconds, and the closest star forming regions are $\gtrsim 100$ pc away, the smallest structures ALMA can resolve are a few AU in size ($\approx 5$ AU in the HL Tau image).  If we assume that the maximum orbital radius of planets is $\sim 100 AU$, $\Delta a / a_1 \gtrsim 10^{-1}$.  In the case of HL Tau, the minimum separation between the putative planets is $\Delta a / a_1 \approx 0.2$.  If we assume most observed stars will be less massive than the Sun ($M_\odot$), then planets in resolvable gaps (beyond a few AU from the star) will have orbital periods $\gtrsim 10$ yrs.  Assumption (\ref{M}) restricts us roughly to planetary masses.  At about this threshold, corrections to the below results from terms that are higher order in the planet-star mass ratio become important.  

We additionally assume that the observed system's age is at most a few Myr old, as this is the median timescale on which gas disks disperse \citep[e.g.][]{Haisch01, Cloutier14}.  We note that all our maximum masses are still valid over longer timescales, they just become less relevant.  On Gyr timescales, a range of more subtle secular and chaotic effects can destabilize progressively lower-mass systems \citep[see, e.g.][]{Lithwick11, Wu11, Batygin15}.  The short timescales in question ($\lesssim 10^5$ orbits) allow us to provide particularly simple estimates.  

Finally, by integrating the current system forward in time, we are implicitly assuming that the current state is representative of its past (or at least migrated into the current state over a timescale comparable to the system's age).  In particular, we assume planets did not recently jump to their current positions.  This is reasonable in the HL Tau system where the orbital periods of the outer planets are $\sim 10^3$ yrs, since simulations of gap opening in the dust component of gaseous disks \citep{Fouchet10} show that the timescale for the planet to clear a gap is $10-100$ orbits.

\section{Instabilities in non-resonant systems} \label{inst}
In this section, we begin by considering the simpler case of a disk-free system, assuming that resonances between planets are not modifying the dynamics.  In subsequent sections, we then sequentially incorporate additional layers of complexity and investigate how they modify our simple estimates.

Much work has been done on the minimum separation between planetary orbits that is required for stability, as a function of mass.  We first discuss the case with two planets orbiting a central star (i.e., the three-body problem), which exhibits qualitatively different behavior from the many-body problem.  

\subsection{Two planets} \label{2pl}
Despite the three-body problem notoriously not possessing enough constants of motion to be integrable, \cite{Marchal82} and \cite{Milani83} showed that, for a large enough combination of total energy and angular momentum, the motion is constrained in such a way that close encounters can never occur (i.e., the system is Hill stable).  For initially circular and coplanar orbits, \cite{Gladman93} translated this stability criterion to a separation of $\Delta a > 3.46 R_H$, where $R_H$ is their mutual Hill radius
\begin{equation} \label{rh}
R_H \equiv a_1 \Bigg( \frac{M_1 + M_2}{3 M_\star} \Bigg)^{1/3}.
\end{equation}

Planets separated by less than 3.5 $R_H$ destabilize on the timescale of conjunctions  $t_\text{conj}$(where planets arrive at the same longitude and close encounters occur).  For ALMA systems, this corresponds to $t_\text{conj} \lesssim 10$ orbital periods of the innermost detected planet\footnote{$t_\text{conj} = 2\pi / \Delta n$ where $\Delta n$ is the difference in the bodies' mean motions $n = 2\pi / P$.  For $\Delta n / n_1 \ll 1$, Kepler's 3rd law yields $t_\text{conj} \approx (\Delta a / a_1)^{-1}$ orbital periods.  For ALMA systems, $\Delta a / a_1 \gtrsim 0.1$.}---much shorter than the disk's lifetime.

In most two-planet cases, separation greater than 3.5 $R_H$ (and thus avoidance of close encounters), is a sufficient condition for stability \citep{Barnes06}.  It is nevertheless possible to more slowly diffuse through chaotic regions of phase space and lose planets \citep[e.g.,][]{Veras13}; however, this is unlikely to play an important role for systems probed by ALMA.  The chaotic region where this is possible corresponds closely to the Hill-unstable region, and only extends above 3.5 $R_H$ for planets with $M \lesssim 10 (M_\star/M_\odot)$ Earth masses ($M_\oplus$)  \citep{Deck13}.  These masses are smaller than those generally found capable of producing perturbations on the dust and gas distributions in protoplanetary disks that would be observable by ALMA, e.g., \citealt{Dong14}.  Additionally, given our range of orbital periods (Eq.\:\ref{P}), \cite{Veras13} find that such evolution is generally slower than the lifetimes of typical protoplanetary disks \citep{Haisch01}.

Therefore, for (non-resonant) two-planet systems observed by ALMA, we conclude that $\Delta a > 3.5 R_H$ furnishes a robust criterion for stability.  For an observed separation, and assuming equal-mass planets, one can invert the equation to obtain the maximum mass for stability $M_\text{crit}$,
\begin{equation} \label{2planet}
M < M_\text{crit} = 12 \Bigg(\frac{M_\star}{M_\odot}\Bigg) \Bigg( \frac{\Delta a / a}{0.1}\Bigg)^3 M_\oplus
\end{equation}
For unequal mass planets, the condition is simply that the sum of the masses be less than twice $M_\text{crit}$.

\subsection{Three or more planets} \label{3pl}

The addition of a third planet expands the degrees of freedom available to the system, such that conservation of the total angular momentum and energy can no longer guarantee the avoidance of close encounters for initial separations beyond 3.5 $R_H$.  A detailed theoretical understanding of stability in systems with three or more planets is currently lacking; however, numerical simulations furnish useful results.  

\cite{Chambers96} found that systems of three terrestrial-mass planets, spaced by the same number of Hill radii from one another, all go unstable, at least for spacings $\Delta a / R_H < 10$.  However, they found that the timescale to instability scales exponentially with $\Delta a / R_H$.  Similar behavior was also observed for giant planets by \cite{Marzari02} and \cite{Chatterjee08}, though the scalings quantitatively differ.  Nevertheless, the above studies agree that for separations $\lesssim 4 R_H$, instability sets in after less than $\sim 10^3$ orbits.  This translates to the conservative stability threshold\footnote{Our definition of the Hill radius differs from that used in the studies above.  We chose the definition given in Eq.\:\ref{rh} because it's the one appropriate in the analytic limit of Eq.\:\ref{2planet}.  In any case, the discrepancies are small, and always act in the direction of making our limit (Eq.\:\ref{3planet}) even more conservative.}
\begin{equation} \label{3planet}
M \lesssim M_\text{crit} = 8 \Bigg(\frac{M_\star}{M_\odot}\Bigg) \Bigg( \frac{\Delta a / a}{0.1}\Bigg)^3 M_\oplus.
\end{equation}
Despite this estimate not being analytic like Eq.\:\ref{2planet}, Eq.\:\ref{3planet} should represent a maximum mass in the sense that it applies to the idealized case of planets equally separated in $\Delta a / a$.  One would expect that as one separated the third planet from the other two, one would recover the 2-planet condition (probably not formally, but certainly over the relevant timescales).  \cite{Marzari14} recently numerically studied such equal-mass but unequally spaced 3-planet systems.  His results suggest that when the separation (in $\Delta a / a$) of the third planet is $\approx 3-4$ times larger than that between the tighter pair, one roughly recovers the two-planet result.  Thus, in a general system, if one identifies the minimum $\Delta a / a$, Eqs.\:\ref{2planet} and \ref{3planet} bracket $M_\text{crit}$ in the limits of an isolated pair of nearby planets (a hierarchical system) and of a chain of equally spaced (in $\Delta a / a$) planets, respectively. 

\subsection{The effects of eccentricity damping} \label{edamping}
Protoplanetary disks will additionally perturb planetary orbits \citep{Papa06}, and this will modify the results of the previous two sections.  These interactions are complicated---they depend on the detailed balance between the gravitational effects of gas at resonant locations with the planet, while the gas densities at these locations depend on the disk's thermal structure and are in turn modified by the resonant torques.  As a result, the planetary mass required to open a gap in the disk \citep[e.g.][]{Bryden00, Crida06}, and the associated timescales on which planetary orbital elements evolve, are uncertain.  Rather than argue for specific scenarios or timescales, we choose simple parametrizations for how the orbital elements will evolve, and consider a wide range of timescales.  One can then use our results to set dynamical constraints by selecting the appropriate timescale under a particular scenario, e.g., a planet that opens a deep gap, a planet that opens a gap in the dust, but not the gas, etc.

We begin by considering the effects of eccentricity damping by the disk \citep{Goldreich80, Arty92, Arty93}.\footnote{We do not consider the possibility of eccentricity excitation by the disk \citep{Goldreich03, Tsang14}, since (at least in closely packed systems) this would rapidly lead to instability.}  Following \cite{Papa00} and \cite{Teyssandier14}, we implement eccentricity damping by including an additional force in our N-body simulations\footnote{A library for parametrized migration, damping of eccentricities and inclinations, and disk-induced precession can be found at \url{https://github.com/dtamayo/rebound/tree/migration} (see the readme on that page).  It can be used to call various REBOUND integrators from within python.},
\begin{equation}
\vec{{\bf F_e}} = - \frac{2}{\tau_e} \dot{r}\:{\bf \hat{r}},
\end{equation}
where $\vec{{\bf F_e}}$ is the eccentricity-damping force vector acting on a planet, $\dot{r}$ is the radial component of the velocity, and ${\bf \hat{r}}$ is the unit vector pointing outward in the radial direction.  This yields an exponential damping of eccentricities on a timescale $\tau_e$.  This assumes that eccentricity is damped at constant angular momentum.

We now qualitatively assess the effect of eccentricity damping on the stability of planetary systems.  As argued above, if a pair of planets is separated by less than $3.5 R_H$, the system will destabilize on a timescale $\lesssim 10$ orbital periods.  Therefore, unless the eccentricity damps on orbital timescales (and objects so tightly coupled to the gas seem unlikely to sculpt the disk into an observable feature), Eq. \:\ref{2planet} furnishes a stringent upper limit on the planetary masses, independent of the disk's properties and/or the depth of the gap the planet opens.  For separations beyond $3.5 R_H$, the gas-free orbital evolution is more gradual; we would expect that if the eccentricity damping timescale is shorter than the instability timescale (which scales exponentially with $\Delta a / R_H$, Sec.\:\ref{3pl}), the disk will suppress instability.

Thus, we conclude that even in the case of three planets equally spaced in $\Delta a/a$ (Sec.\:\ref{3pl}), one should approach the two-planet result (Eq.\:\ref{2planet}) in the limit of strong eccentricity damping.  We show this in our simulations of the HL Tau system in the next section.

\subsection{Application to HL Tau} \label{hltau}
\begin{figure}[!ht]
\includegraphics[width=\columnwidth]{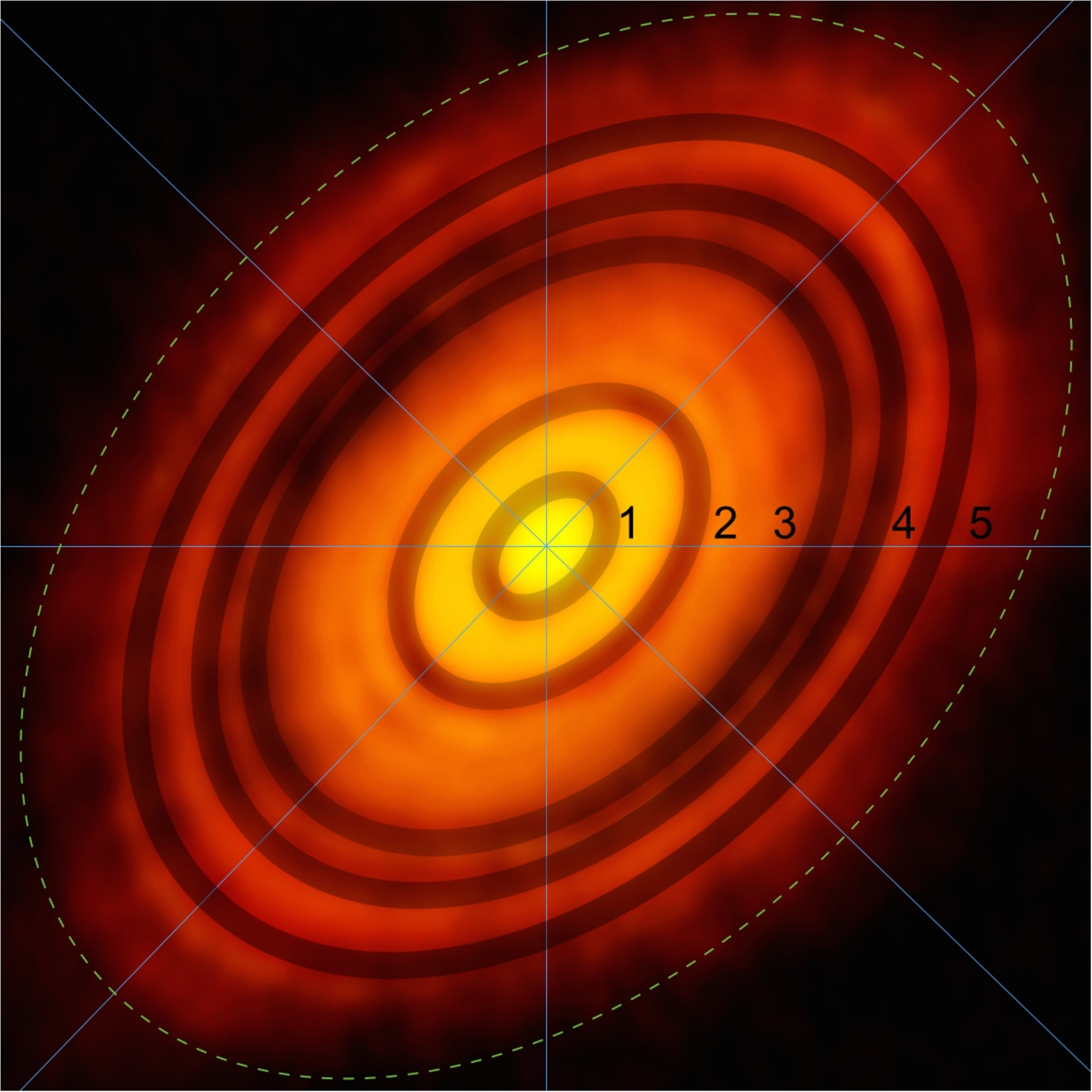}
\caption{\label{hltauimg} Publicly released continuum 233 GHz image of HL Tau taken during ALMA's science verification process.  The gaps we identified and used in our simulations (see Appendix \ref{Amaury}) are displayed. Blue lines aid centering the ellipses along the major and minor axes. The dashed ellipse shows our assumed outer limit of the disc, which we used to set our distance scale (Appendix \ref{Amaury}).  Taken from \url{http://www.eso.org/public/news/eso1436/}. }
\end{figure}

We now apply the above constraints to the HL Tau system and compare our estimates to numerical integrations.  Not knowing the stretch in the publicly released image, we chose to focus on the most significant gaps, labeled in Fig.\:\ref{hltauimg}.  We discuss in Sec.\:\ref{depic} how additional planets change our results.  We describe how we determined the orbital radii corresponding to each gap, and provide a table of values in Appendix \ref{Amaury}.  Additionally, to limit the parameter space, we assume that putative planets all have equal mass.

Our nominal radii of 13.6, 33.3, 65.1, 77.3 and 93.0 AU (table \ref{gaps}) correspond to relative spacings between adjacent gaps $(a_{i+1} - a_i)/a_i$ of 1.4, 1.0, 0.19, and 0.20.  Since the smallest separation is shared by two adjacent pairs, we are in the limit of Eq.\:\ref{3planet}.  Additionally, as discussed above, since the inner two putative planets are $\gtrsim 5$ times more separated than the outer three, the former are unlikely to affect the dynamics.  Indeed, though the analyses described below include all five planets, at our grid resolution, we obtained indistinguishable results for the system's stability with only the three outer planets.  In all our integrations, we adopt $M_\star = 0.55 M_\odot$, estimated by \cite{Beckwith90} using protostellar evolution tracks on luminosity-temperature plots.

We therefore estimate from Eq.\:\ref{3planet} that the planets must have $M \lesssim 30 M_\oplus$.  Additionally, in the limit of strong eccentricity damping (with a timescale somewhat longer than the conjunction time of $\approx (0.2)^{-1}$ orbits $\sim$ a few thousand yrs), we expect that the slower 3-body instability will be suppressed, and that we should recover the 2-planet limit preventing close encounters (Eq.\:\ref{2planet}).  This corresponds to $M \lesssim 45 M_\oplus$.  

For our numerical simulations, we used the high-order integrator IAS15 \citep{Rein15}, which is part of the REBOUND package \citep{Rein12}.  We define instability as when a planet's radial excursions become greater than 5 AU, either through semimajor axis variations, or through radial variations ($a e$) induced by an excited orbital eccentricity.  Like previous studies \citep[e.g.,][]{Gladman93}, we find our results are not sensitive to exactly how we define our criterion.

\begin{figure}[!ht]
\includegraphics[width=\columnwidth]{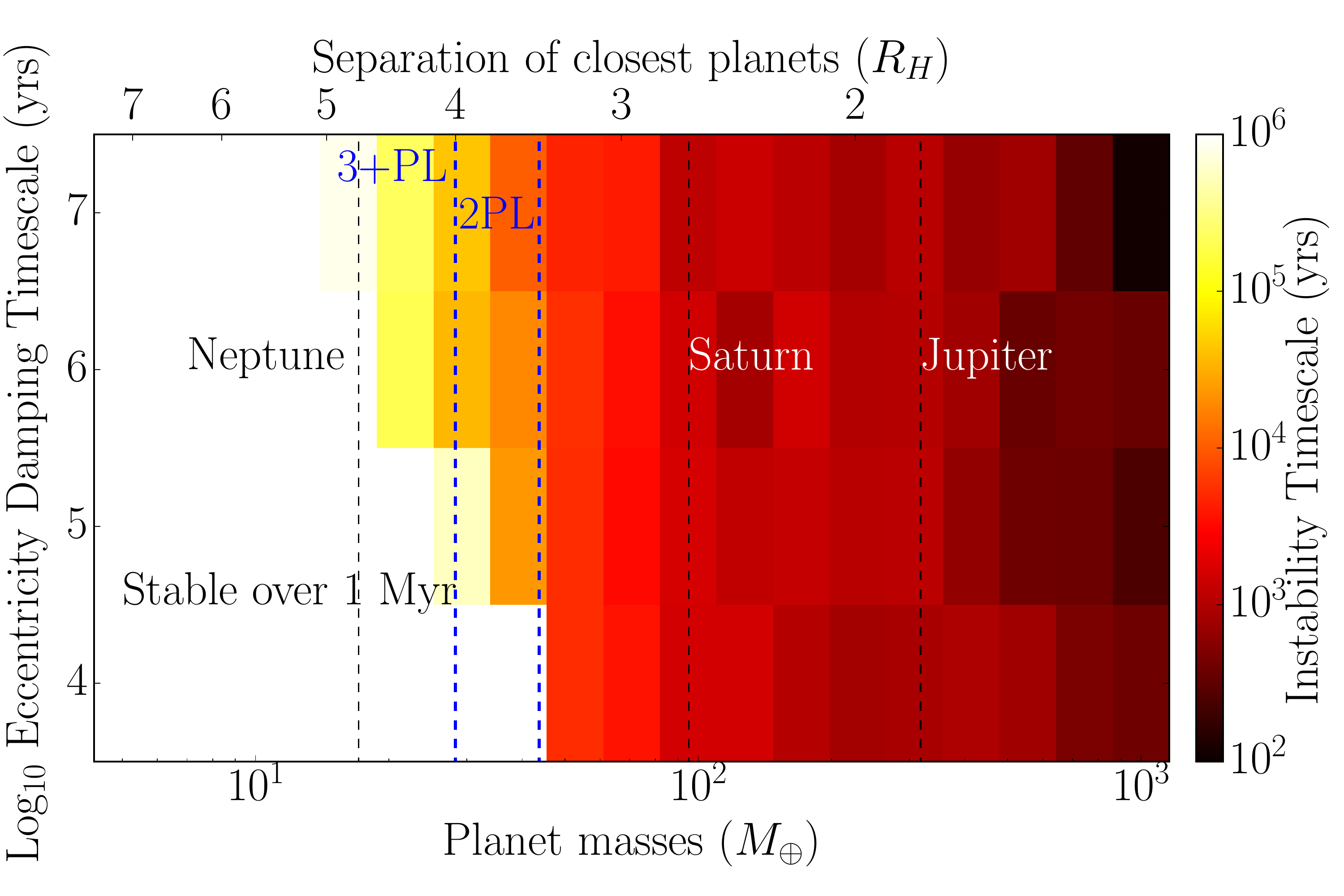}
\caption{\label{5planet} Time to instability for our nominal 5-planet case, as a function of planet mass, and eccentricity damping timescale.  The top axis shows the corresponding separation between the closest planets in units of Hill radii.  Each grid point is the median timescale to instability from 24 numerical integrations (see text).  The vertical blue lines labeled 2 PL and 3+ PL correspond to the predictions expected in the strong eccentricity-damping and dissipation-free limits at 3.5 and 4 $R_H$, respectively (see Eqs.\:\ref{2planet} and \ref{3planet}).}
\end{figure}

In Fig.\:\ref{5planet} we show the time to instability as the masses of the planets and the eccentricity damping timescale are varied.  Each grid point corresponds to 24 numerical integrations lasting $10^6$ yrs (the upper-limit for the star's age), where the particles were initialized at random longitudes on effectively circular and coplanar orbits ($e = 10^{-8}, i = 10^{-8}$ rad $\sim 10^{-6}$ deg, where $i$ is the orbital inclination).  The color in each grid point represents the median of the time to instability (or $10^6$ yrs for runs that remained stable) across the 24 simulations.  We also tried runs with 96 simulations per grid point and obtained visually indistinguishable results.  Along the top of Fig.\:\ref{5planet}, we use the mass to express the separation between the closest planets in units of mutual Hill radii (see Sec.\:\ref{inst}).  

We see that the system is somewhat more unstable than our estimates would suggest.  While this is not necessarily cause for alarm (Eq.\:\ref{3planet} provides a maximum mass---lower masses may still be unstable), the main reason for the discrepancy is that the outer two planets are near the 4:3 resonance (their period ratios are $\approx 1.32$).  In the next section, we discuss how such resonances modify the simple picture painted above. Nevertheless, we see the effect expected from the previous section that for decreasing $\tau_e$, one approaches the 3.5 $R_H$ threshold.

\begin{figure}[!ht]
\includegraphics[width=\columnwidth]{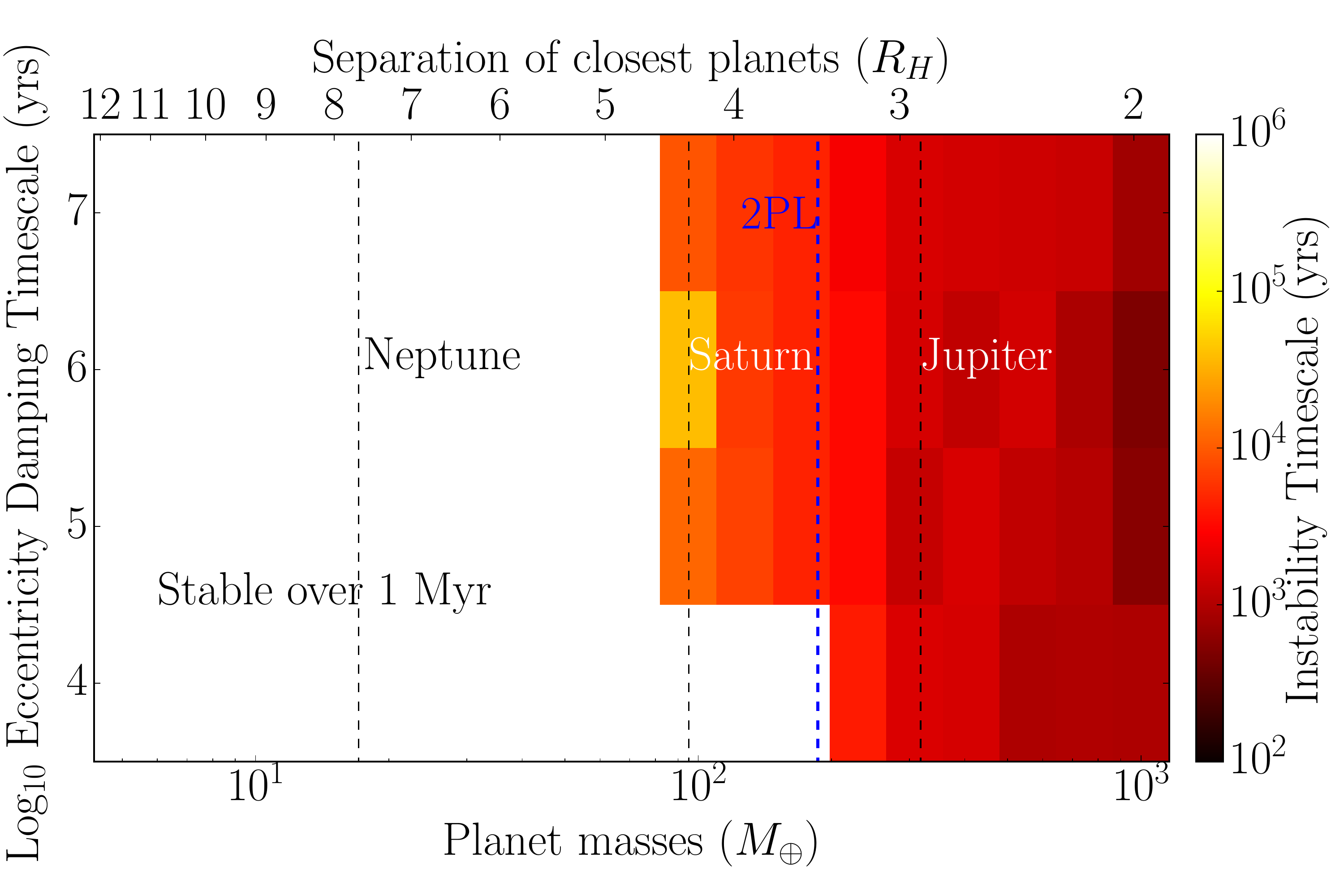}
\caption{\label{4planet} Time to instability for a 4-planet case where gaps 3 and 4 are the result of a single planet with co-orbital material in between.  Otherwise same as Fig.\:\ref{5planet}.}
\end{figure}

Because of the sensitive dependence of stability on separation, one simple possibility that would render the system substantially more stable is if not all gaps correspond to planets.  For example, gaps 3 and 4 may not represent two separate planets, but rather a single one, with co-orbital material on horseshoe orbits in between.  We therefore also consider a system with four planets at 13.6, 33.3, 71.2 (the average of gaps 3 and 4) and 93.0 AU (Fig.\:\ref{4planet}).  This corresponds to $\Delta a/a = 1.4, 1.1$ and 0.31.  Given the single close pair of planets, and inner two planets at much wider separation, we would expect the stability limit to lie close to 195 $M_\oplus$ (Eq.\:\ref{2planet}); however, the two outer planets' period ratio is now $\approx 1.49$, near the 3:2 resonance, which pushes the stability limit to lower masses (because of our $e\approx0$ initial conditions---see Sec.\:\ref{res}).  

\subsection{Dependence on initial conditions} \label{depic}
Before investigating the role of resonances, we briefly justify our choices of initial conditions, and discuss how they affect our results.  We first address our effectively co-planar geometry.  One would only expect inclinations to matter on these timescales if they are capable of mitigating close encounters.  This should roughly occur when the maximum height reached on an inclined orbit $a \sin{i}$ becomes comparable to the separation between planets $\Delta a$, since beyond this inclination, strong encounters would only occur when conjunctions happened to line up with one of the nodes where the two planets' orbital planes cross one another.  However, we argued (Eq.\:\ref{da}) that resolution is likely to limit ALMA to finding separations $\Delta a/a \gtrsim 0.1$.  This corresponds to $i \gtrsim 10$ deg, which is high if the disk effectively damps inclinations \citep[e.g.,][]{Papa00}.  In any case, our brief experimentations with such large inclinations suggest that planets avoid ejections through the additional constraint that close encounters must occur near the nodes; however, they tend to evolve onto crossing orbits that would not produce the observed concentric rings.  

Redoing the simulations in Fig.\:\ref{5planet}, but now drawing inclinations from Rayleigh distributions with $i = 10^{-4}$ and $10^{-2}$ rad ($\approx 0.006$ and $0.6$ deg), both with and without inclination damping, we find visually indistinguishable results from Fig.\:\ref{5planet}.  This motivated our choice for the nominal scenario presented in Sec.\:\ref{hltau} with essentially coplanar initial conditions, ignoring inclination damping.  We briefly note that, while \cite{Barnes15} found that planets in resonances with small inclinations could chaotically evolve onto extremely eccentric orbits that would certainly destabilize a tightly packed system, this occurs on much longer timescales than we are considering.  

As for the eccentricities, one might expect that the initial values are not significant as long as they are much smaller than the eccentricity required for crossing orbits $e_\text{cross} \sim \Delta a/a$, which is approximately 0.2 in our HL Tau simulations.  Indeed, if we redo the analysis in Fig.\:\ref{5planet} (which has initially circular orbits), drawing the various eccentricities from Rayleigh distributions\footnote{with the Rayleigh probability density function $f$ defined through $f(e;\sigma) = x \text{exp}[-x^2 / (2\sigma^2) ] / \sigma^2$, where $\sigma$ is the scale parameter.} with scale parameters of $10^{-4}$ and $10^{-2}$ (and the longitudes and pericenters randomized), we obtain visually indistinguishable results from Fig.\:\ref{5planet}.  However, as we discuss in the next section, this is likely not the appropriate way to choose initial conditions when planets are near resonance.

We also point out that our assumed mass of $0.55 M_\odot$ \citep{Beckwith90} is uncertain.  Deviations from this value would affect our result; however, solving for a separation of a certain number of Hill radii (Eqs.\:\ref{2planet} and \ref{3planet}) yields the mass ratio between star and planet.  Thus, our values (and errors) scale directly with the star mass.  The secondary effect that the orbital periods will change is not as significant, given the exponential rise in instability timescale with separation.

Finally, we comment on the inclusion of additional planets to our nominal case.  There may be two faint gaps between gaps 2 and 3 in our nominal model (Fig.\:\ref{hltauimg}), as well as a possible gap near the outer edge.  If we simulate as above a system of these 8 planets (with orbital radii of 43.4, 53.4 and 107.2), we obtain qualitatively similar results to Fig.\:\ref{5planet}, though shifted over one bin to lower masses at our level of resolution.  The small discrepancy reflects the fact that the additional planets do not introduce much smaller $\Delta a /a$ values (though the outermost two planets have a somewhat smaller $\Delta a / a$ of 0.15).  However, one important difference is that, with the increased number of planets, it becomes harder to find orbital radii within the errors that are not close to {\it some} first-order mean motion resonance (MMR) (while in the next section we easily find such configurations for the nominal five-planet case).  Resonances are therefore important in an eight-planet scenario (see next section).  We note that this does not necessarily imply an eight-planet scenario requires lower masses---the reduction in stability is a product of our choice of initial conditions (Sec.\:\ref{res}).
\section{Resonant effects} \label{res}
Resonances can be perplexing in that they sometimes provide great stability (e.g., the 2:3 resonance between Pluto and Neptune that has prevented collisions over several Gyr despite their crossing orbits), while in others cases they are destabilizing (e.g., the Kirkwood gaps at resonant locations with Jupiter in the asteroid belt).  We now briefly review the relevant dynamics in order to identify the relevant regimes and provide a lens through which to interpret numerical investigations.  

\subsection{Background} \label{back}

On the short timescales we are interested in, systems destabilize through close encounters at conjunctions.  Because resonances modify both the planets' orbital eccentricities, as well as where the conjunctions occur, they can have a strong effect.  As we discuss below, whether these changes help or hinder stability depends on initial conditions, so these must be judiciously chosen.  For studies of stability near the 3:1 and 2:1 resonances on longer timescales without a gas disk, see \cite{Marzari05, Marzari06}.

It is well known that, near an isolated MMR between two planets on nearly circular and co-planar orbits, one can obtain an approximate one-degree-of-freedom Hamiltonian that can be analytically integrated \citep[e.g.,][]{Message66, Peale86}.  Sacrificing precision for simple estimates, we draw intuition from the case of a massless particle perturbed by a massive planet on an exterior, coplanar and circular orbit.\footnote{For a particularly clear (but involved!) treatment of the more general case with two massive bodies on elliptical orbits, see \cite{Deck13}.  They show that one obtains the same dynamics as in the test-particle case for first-order MMRs, when expressed in terms of a generalized eccentricity that is a weighted sum of the two planets' orbital eccentricities.  Our simple treatment thus provides rough but qualitatively accurate estimates.}  We additionally specialize to the case of first-order mean motion resonances (with period ratios of the form $j+1:j$), which dominate when the eccentricities are small \citep[see][Chap. 8]{Murray99}.

In this case, one obtains a solution in terms of the eccentricity of the test particle $e_1$ (the outer orbit is circular and fixed) and the resonant angle $\phi = j \lambda_1 - (j+1) \lambda_2 - \varpi_1$, where $\lambda$ and $\varpi$ correspond to the particles' mean longitudes and their longitudes of pericenter, respectively.  Physically, one can think of the resonant angle as tracing where conjunctions will occur, as when the planets line up ($\lambda_1 = \lambda_2$), $\phi = \lambda - \varpi_1$, i.e., $\phi$ is the longitude of conjunction, measured from pericenter.  This is an important angle since, for example, if conjunctions always happen at pericenter, this maximizes the distance between particles; as a result, $\phi = 0$ is the stable configuration for first-order MMRs.  Exact resonance occurs at semimajor axes that render $\phi$ stationary, i.e.,
\begin{equation} \label{phidot}
\dot{\phi} = j n_1 - (j+1) n_2 - \dot{\varpi_1} = 0,
\end{equation}
where $n$ represents the mean motion $2\pi/P$.  This can be trivially rearranged to give
\begin{equation} \label{rescond}
\frac{P_2}{P_1} = \frac{j+1}{j} - \frac{\dot{\varpi_1}}{jn_2}.
\end{equation}
Generally, $\varpi_1$ will precess slowly compared to the planets' orbital rates, so exact resonance approximately corresponds to the intuitive separation where $P_2/P_1 = (j+1)/ j$; however, we will see below that this is not always the case.

\begin{figure}[!ht]
\includegraphics[width=\columnwidth]{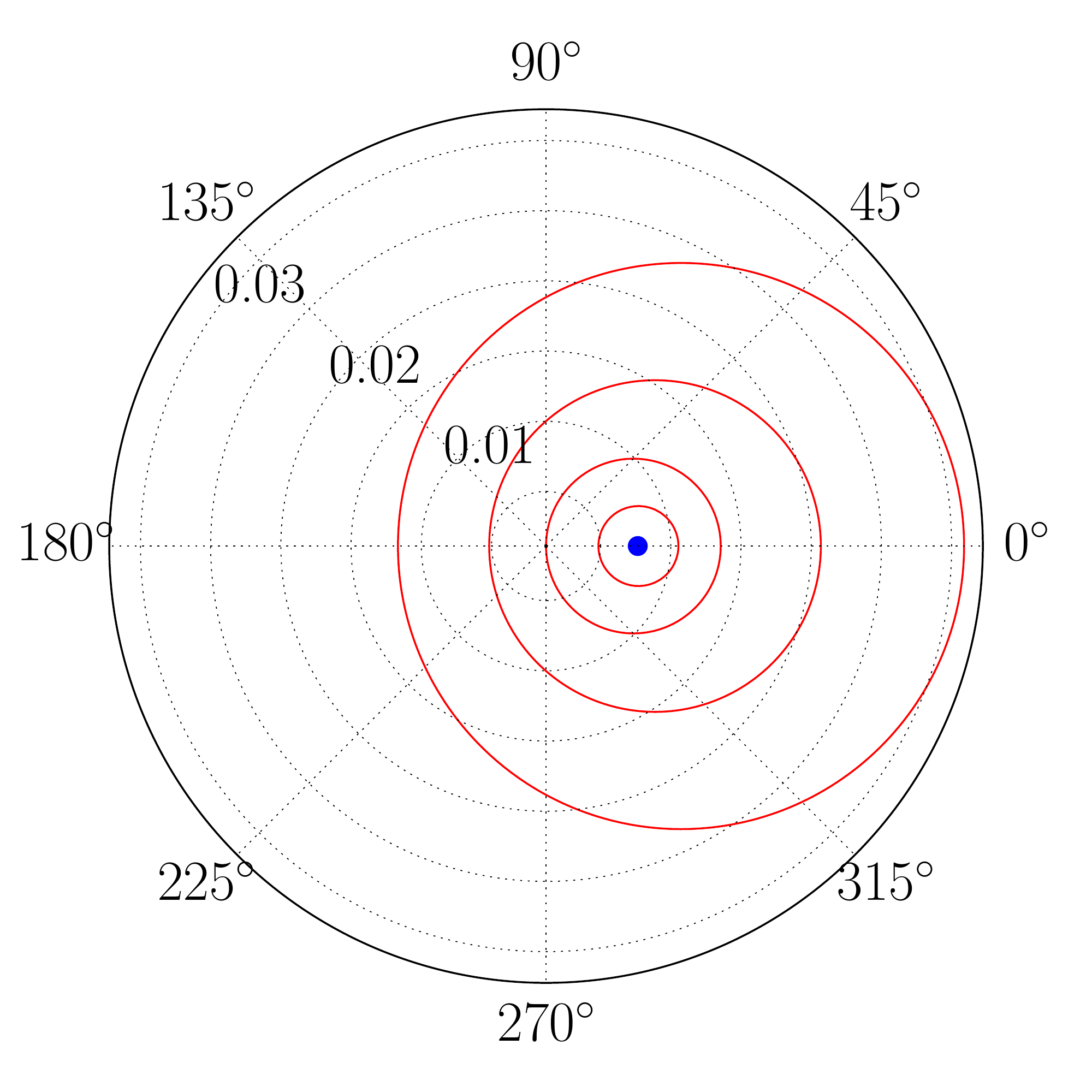}
\caption{\label{resfig} Polar plot of $e$ and $\phi$ for a test particle near resonance with an outer perturber on an exterior circular orbit.  The fixed point's location at $e_f\approx 0.008$ (blue dot), is set by Eq.\:\ref{ef}, where we have used $\mu = 10^{-4}$ and $\Delta = 0.01$ (Eq.\:\ref{delta}).  Values of $(e,\phi)$ for systems with four different initial conditions trace out the red solid curves.  As the point describing the system moves along one of these curves, $e$ and $\phi$ vary (the point lies at different azimuths and intersects different dashed circles of constant eccentricity).} As long as the initial eccentricity $e_0$ is not $\gg e_f$ (in which case the system traces approximately traces out a circle at constant $e_0$), the eccentricity variations are comparable to $e_f$.
\end{figure}

A useful way to visualize the dynamics is as a polar plot of $e_1$ and $\phi$; see Fig.\:\ref{resfig}.  A point on this plot specifies the state of the system at some time, with the eccentricity given by the point's radial distance from the origin (the dashed gray circles of constant eccentricity are labeled at $e=0.01,0.02,0.03$), and $\phi$ by the angle from the positive $x$ axis.  For a given planet separation that yields periods approximately satisfying Eq.\:\ref{rescond}, one finds that along $\phi = 0$, there is a stable equilibrium for a particular value of the eccentricity $e_f$, referred to as the forced eccentricity.  For initial conditions slightly displaced from the equilibrium point, the system traces out a closed contour around this fixed point, corresponding to oscillations in time for both the eccentricity and $\phi$.  

On short timescales, resonances can either help or hurt stability, depending on initial conditions.  Superposed on the smooth evolution shown in Fig.\:\ref{resfig} are discrete kicks at conjunctions.  If the system starts sufficiently far from the equilibrium, the potentially large rises in eccentricity, together with the variations in conjunction longitude, can lead to close encounters that quickly destabilize the system.  Conversely, for initial conditions close to equilibrium, conjunctions always happen at the most stable longitude, minimizing the strength of the kicks incurred at conjunctions.

The equilibrium location is set by the planet-pair's proximity to exact resonance.  When the planets' periods are far from resonance, the equilibrium configuration is for both orbits to be circular.  As one brings the semimajor axes toward the resonant separation, the equilibrium eccentricity shifts away from the origin toward the right along the $x$ axis.  This so-called forced eccentricity is given by 
\begin{equation} \label{ef}
e_f \approx \frac{\mu}{\Delta},
\end{equation}
where we have omitted a coefficient of order unity ($\approx 0.8$ in our case, see Eq. 11 in \citealt{Goldreich14}), $\mu$ is the ratio of the perturber's mass to that of the central body, and $\Delta$ is a dimensionless measure of the proximity to resonance \citep{Lithwick12}\footnote{We have borrowed the intuitive variable $\Delta$ from \cite{Lithwick12}, but redefined it in a way that renders Eq.\:\ref{ef} particularly simple.  Our $\Delta$ thus varies trivially from theirs, and agrees for $\Delta \ll 1$.}
\begin{equation} \label{delta}
\Delta \equiv \frac{P_1}{P_2}\Bigg(\frac{P_2}{P_1} - \frac{j+1}{j}\Bigg) = 1 - \frac{j+1}{j}\Bigg(\frac{a_1}{a_2}\Bigg)^{3/2},
\end{equation}
where the first quantity in parentheses approximately approaches zero at resonance.  When planets are more widely separated than the resonant separation ratio, $\Delta > 0$, and when they are closer together, $\Delta < 0$.\footnote{Equation \ref{ef} unphysically diverges at $\Delta = 0$.  This is due to a truncation of the equations of motion at leading order in the small quantities $e$ and $\mu$.  This is not a limitation since, as discussed below, dissipation prevents planets from approaching exact resonance.}  We note that when $\Delta < 0$, Eq.\:\ref{ef} implies $e_f < 0$.  This is the correct result for one of the fixed points if one considers a negative eccentricity to correspond to $\phi = \pi$ rather than $\phi = 0$.  

For rough estimates close to resonance (the region of interest), one can write the approximate forms
\begin{eqnarray}
\Delta &\sim& \frac{P_2}{P_1} - \frac{j+1}{j}, \\ \label{pdelta}
&\sim& \frac{a_2}{a_1} - \Bigg(\frac{a_2}{a_1}\Bigg)_\text{res}, \label{adelta}
\end{eqnarray}
where $(a_2/a_1)_\text{res}$ is the semimajor axis at nominal resonance $[(j+1)/j]^{2/3}$.  Close to resonance, the variable $\Delta$ is approximately therefore simply the shift of the period or semimajor axis ratio from the nominal resonance value (i.e., Eq.\:\ref{rescond} with $\dot{\varpi} = 0$).  We further note that Eq.\:\ref{adelta} generalizes to the case discussed in Sec.\:\ref{prec} where the resonant locations shift.  In that case, $(a_2/a_1)_\text{res}$ would correspond to the semimajor axis ratio that satisfies Eq.\:\ref{phidot}.

\subsection{When are resonances important?} \label{resimp}
We are now in a position to evaluate the effects of resonances on our analysis in Sec.\:\ref{hltau}.  We found in the previous section that, for a given $\Delta$, the fixed point in the dynamics lies at $e_f \sim \mu/\Delta$ (Eq.\:\ref{ef}).  Assuming the initial eccentricities aren't $\gg e_f$ (see caption to Fig.\:\ref{resfig}), $e_f$ sets the scale for the eccentricities the system will reach---orbits at the fixed point will remain at $e=e_f$, while in the limiting case of initially circular orbits, the eccentricity will approximately vary between zero and $2e_f$ (Fig.\:\ref{resfig}).  

This can explain the discrepancies in the figures of Sec.\:\ref{hltau} from the expected mass thresholds in Eqs.\:\ref{2planet} and \ref{3planet}.  Those simulations started with circular orbits.  If the semimajor axes for a pair of planets lie within some $\Delta$ of resonance, the resultant variations in $e \sim \mu/\Delta$ become an appreciable fraction of the eccentricity required for orbits to cross $e_\text{cross} \approx \Delta a / a$.

In our nominal 5-planet case, we expected to find the mass threshold at $\approx 30 M_\oplus$, or $\mu = M_p/M_\star \approx 2 \times 10^{-4}$.  Additionally, the period ratio between the outermost two planets $P_5/P_4 \approx 1.32$, which is close to the first-order 4:3 MMR (with period ratio 1.33).  From Eq.\:\ref{delta}, $\Delta \approx -0.01$, so we expect eccentricities to reach $e \sim 0.02$ by Eq.\:\ref{ef}.  This is an appreciable fraction ($\sim 10\%$) of the eccentricity required for orbit-crossing $e_\text{cross} \approx \Delta a / a \approx 0.2$, which hastens the onset of instablility.

\begin{figure}[!ht]
\includegraphics[width=\columnwidth]{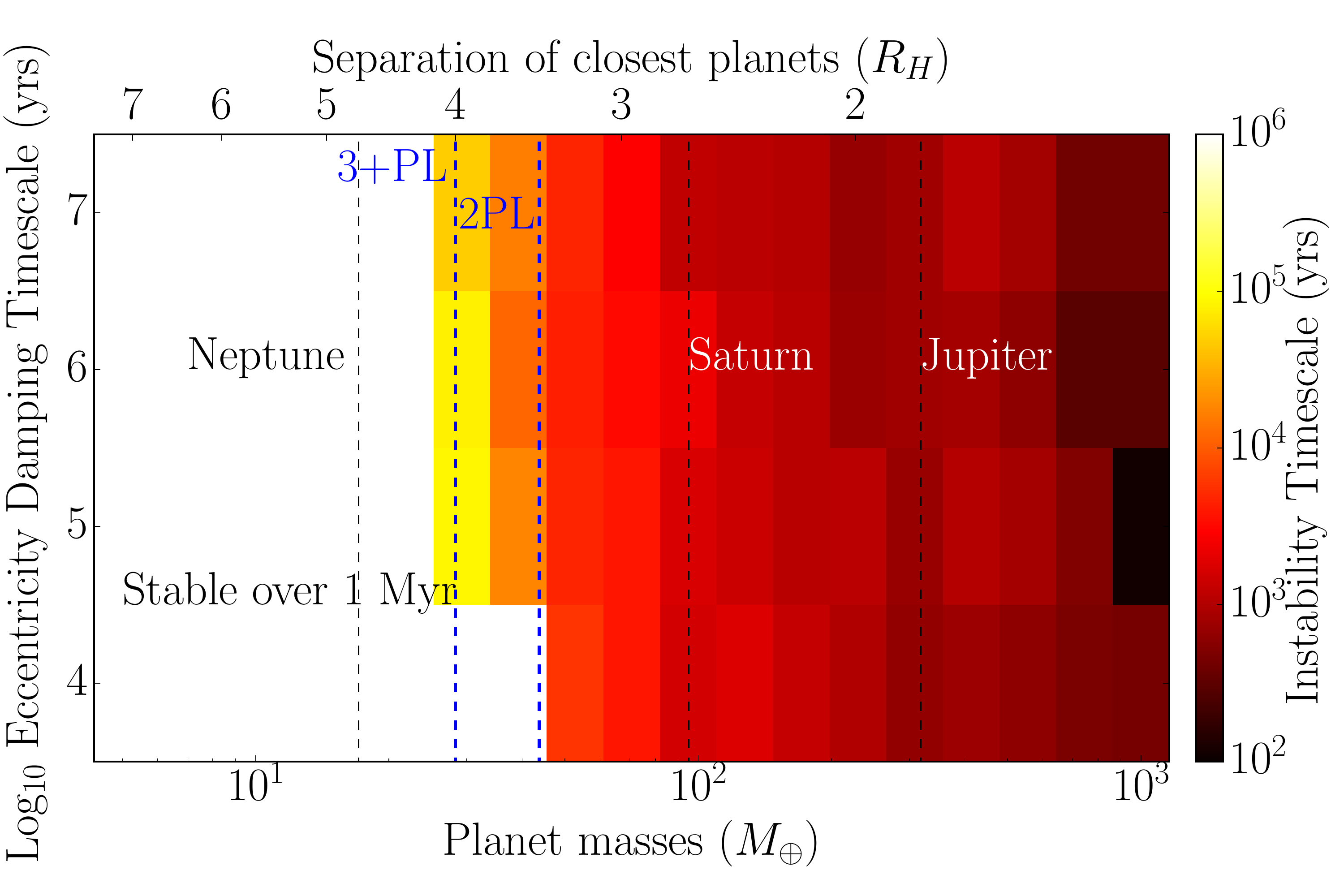}
\caption{\label{5pshift} Time to instability for a non-resonant case (the outermost planet has been shifted inward by 1 AU relative to our nominal setup).  Otherwise same as Fig.\:\ref{5planet}.}
\end{figure}

If we move the system away from resonance by shifting the initial semimajor axis of the outermost planet inward by 1 AU, $\Delta$ triples, correspondingly decreasing the magnitude of eccentricity oscillations.  Performing the same analysis as in Sec.\:\ref{hltau} with this shifted system, we obtain Fig.\:\ref{5pshift}, which matches well the simple expectations from Sec.\:\ref{inst}.\footnote{The extra stability in Fig.\:\ref{5pshift} at $\tau_e = 10^4$ yrs is due to the damping timescale being faster than the slower instability timescale beyond 3.5 $R_H$, as discussed in Sec.\:\ref{edamping}.}  This suggests an approximate empirical threshold for $e_f$ of $\sim 10\%$ of the eccentricity required for orbit-crossing.

To investigate this claim further, we ran a large suite of numerical integrations for two planets in the limits of a massless inner planet and of two equal-mass planets.  The inner planet was placed at the location of gap 3, and the second was placed at various radii near the locations of the major first-order MMRs (2:1, 3:2 and 4:3).  We started the planets on circular orbits, and for a given proximity to a particular resonance $\Delta$, we varied the planet mass to see at what value of $e_f = \mu / \Delta$ (Eq.\ref{ef}) the resonantly induced eccentricity oscillations led to instability within $10^6$ yrs.  As might be expected, this critical $e_f$ varies with $\Delta$ and the particular resonance; however, we find that in all cases, the critical $e_f$ is less than 0.01.  We can thus define (through Eq.\:\ref{ef})
\begin{equation} \label{deltacrit}
\Delta_\text{crit} \equiv 100 \mu.
\end{equation}
This provides a useful rule-of-thumb:  For planets near the 2:1,3:2, and 4:3, if $\Delta > \Delta_\text{crit}$, then the resonances do not significantly affect stability, and Eqs.\:\ref{2planet} and \ref{3planet} should provide a robust mass limit for stability.  This is also the limit in which ``naive" initial conditions of circular orbits are appropriate for a numerical stability analysis in an eccentricity-damping disk.  Conversely, if $\Delta < \Delta_\text{crit}$, stability will be more sensitive to initial conditions and a more detailed analytical or numerical analysis is required (see Sec.\:\ref{upper}).  It is important to note that this is a statement about stability over short timescales of $10^3-10^4$ orbits, and that, particularly for the 4:3, one should ensure that the $\Delta$ values one considers does not place the system closer to a different first-order MMR.

\begin{figure}[!ht]
\includegraphics[width=\columnwidth]{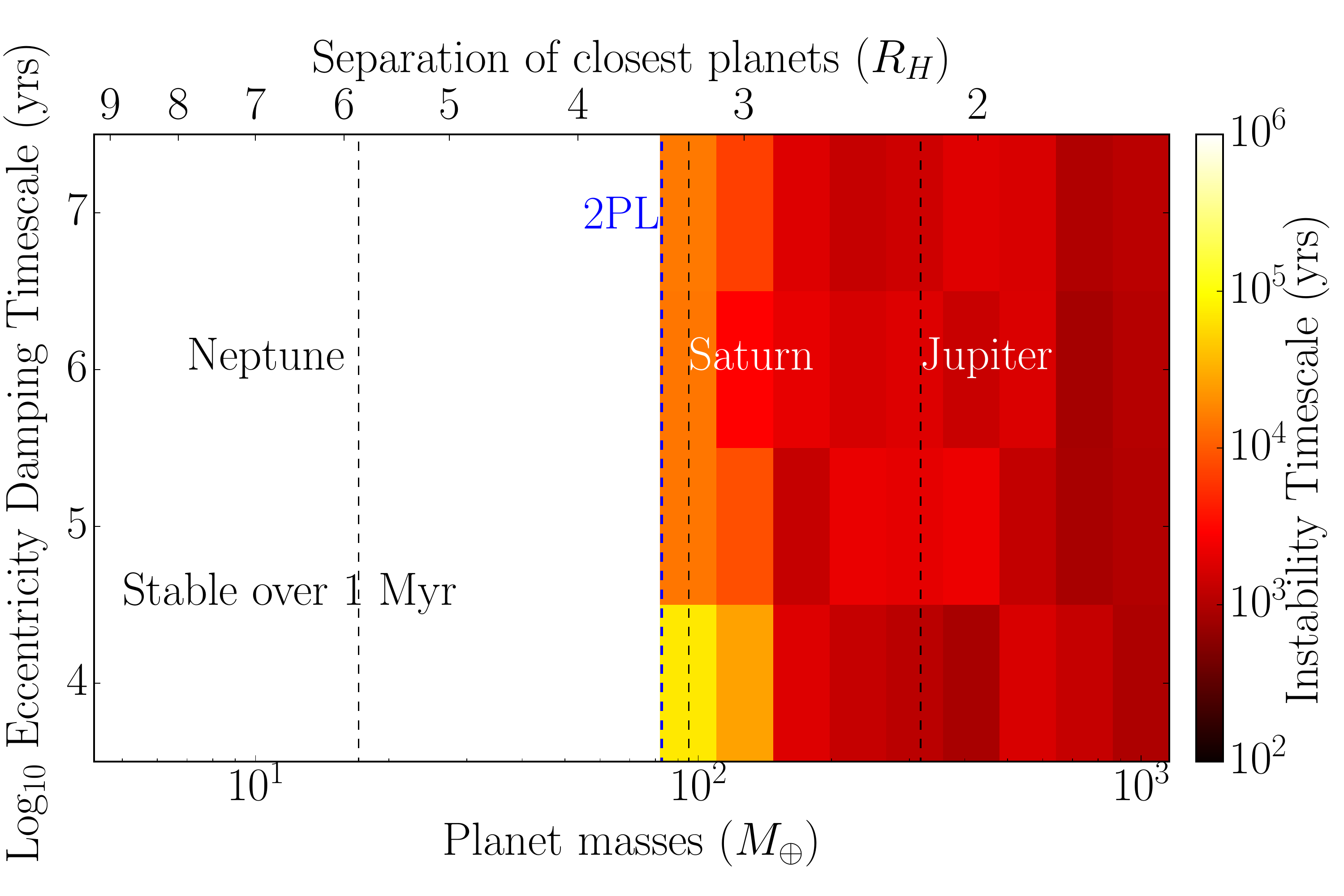}
\caption{\label{4pshift} Time to instability for a non-resonant case of our 4-planet system (with the planet with co-orbital material moved from 71.2 to 75.5 AU).}
\end{figure}

As a second example, in our four-planet case (Fig.\:\ref{4planet}), the expected mass was approximately half a Jupiter mass, or $\mu \approx 10^{-3}$, and $\Delta \approx {\bf 0.005}$.  We therefore expect eccentricitiy variations $\sim 0.1$, which should be very unstable, and helps explain the large deviations from the expected thresholds in Fig.\:\ref{4planet}.  From Eq.\:\ref{deltacrit} we require $\Delta \gtrsim \Delta_\text{crit} = 0.1$.  This can be achieved by shifting the second-to-last planet outward by 4 AU.  Figure \ref{4pshift} shows the result, which matches very closely the analytic threshold from Eq.\:\ref{2planet}.
One should compare the shifts required to move between resonant and non-resonant cases to the observational uncertainties in the orbital radii.  But before doing so, we must consider an important complication.

\subsection{The resonances are not where you think they are} \label{prec}
The locations of resonances in a system are usually accurately determined by simple period ratios, and are thus independent of both the central star's mass and the absolutely size scale of the system.  However, this is only true when pericenter precession rates are slow compared to the orbital rates (see Eq.\:\ref{rescond}).  This is generally a safe assumption, as the pericenter direction is a conserved quantity in the Kepler problem, so small non-keplerian perturbations $F_\text{NK}$ induce a slow precession of order $\varpi/n \sim F_\text{NK}/F_K \ll 1$, where $F_K$ is the dominant gravitational force from the central star.  

However, for young systems, the disk's gravity can present a significant non-Keplerian component.  In the case of HL Tau, \cite{Kwon11} estimate a disk mass $M_D$ of 0.13 $M_\odot$, which represents about one quarter of the central star's mass.  Assuming a razor-thin, axisymmetric disk, this would induce a pericenter precession rate (relative to the mean motion) of $\dot{\varpi}/n \sim M_D/M_\star \sim 0.25$ (Eq.\:\ref{precrate}).  According to Eq.\:\ref{rescond}, this would move the location of the 4:3 resonance by $\sim 10\%$.  This is a significant shift, since we found in the previous section that in our nominal five-planet model, the 4:3 resonance between the outermost two planets was only important within $\Delta \sim 1\%$ of resonance.

We point out, however, that the precession rate derived in Eq.\:\ref{precrate} is very uncertain, as it ignores deviations from a smooth axisymmetric disk.  These can be particularly large in the vicinity of the planet (e.g., any gap or spiral density waves induced by the body), and would provide contributions to the precession rate that are difficult to calculate due to the feedback between the planet and disk, the dependence on disk structure and thermodynamics, etc.  Nevertheless, Eq.\:\ref{precrate} provides the scale on which the precession rates (and therefore the resonance locations) are uncertain.  We also note that this base pericenter precession rate induced by a smooth axisymmetric disk is negative, which means through Eq.\:\ref{rescond} that resonances lie at wider period ratios, and thus at wider separations, than one would expect in the disk-free case.  This means that the outermost two pairs of planets in our nominal five-planet case could plausibly lie in 5:4 resonances, despite their naive period ratios (using Kepler's 3rd law) putting them closer to the 4:3.

We note that in contrast to the simple case considered in Sec.\:\ref{back} where the inner planet is massless, in the general case of two massive bodies, both orbits evolve, and there exists a second resonant argument involving the longitude of pericenter of the outer body.  \cite{Beauge03} \citep[see also][]{Beauge06} find that in the disk-less case, the equilibrium solutions in a secular approximation correspond to the two pericenters precessing at the same rate (apsidal corotation).  An interesting question we leave for future work is whether in a massive disk, the differential pericenter precession rates induced by the disk are strong enough to break the apsidal corotation that is expected from slow resonant capture \citep{Beauge06}.  Additionally, this interaction between planets acting in addition to the disk should affect the precise location of the shifted resonances; however, because the disk-induced precession is so uncertain, we do not pursue this further.

\subsection{Can one observationally constrain whether ALMA planets are in resonance?} \label{errors}
The previous section argued that the locations of resonances are uncertain for planets in a disk by of order the ratio of the disk's mass to the central star.  While this ratio is typically observed to be $\sim 1\%$ \citep[see Fig. 5 in][]{Williams11}, there is a large scatter of $\sim \pm 1$ dex, and one might expect early ALMA observations to target brighter disks.  This suggests that in many systems, even if one could extract exact semimajor axes for each of the planets, one could not determine the proximity to resonance $\Delta$ to within the accuracy needed to tell whether the resonant interactions are strong (Eq.\:\ref{deltacrit}).  

Additionally, there are observational uncertainties in the orbital radii due to the spatial resolution.  From Eq.\:\ref{adelta}, the error in $\Delta$ ($\delta \Delta$) associated with an error $\delta a$ on the semimajor axis is
\begin{equation}
\delta \Delta \approx \frac{3}{2} \sqrt{\Bigg(\frac{\delta a}{a_1}\Bigg)^2 + \Bigg(\frac{\delta a}{a_2}\Bigg)^2} \approx 2 \frac{\delta a}{a_1}.
\end{equation}
In the HL Tau image, the resolution is about 5 AU\footnote{\url{https://public.nrao.edu/news/pressreleases/planet-formation-alma}}, which corresponds to an uncertainty in $\Delta$ of $\approx 0.1$.

Put together, this suggests that one is unlikely to determine whether planets in ALMA systems are resonantly interacting from observed gap radii.  Conversely, it will be similarly difficult to {\it rule out} resonances in tightly-packed systems (all semimajor axis ratios less than 1.4 are within $\approx 10\%$ of a first-order MMR).  However, a promising avenue for observationally breaking this degeneracy is to search for eccentric gaps (see Sec.\:\ref{maxres}).

\section{Upper mass limits in ALMA systems} \label{upper}
Given the above uncertainties, we proceed to estimate the maximum masses across the range of possibilities.  This will correspond to systems that are stabilized near the centers of resonances (Fig.\:\ref{resfig}).  We have already seen that in resonant configurations, different initial conditions can lead to very different outcomes.  This presents a large parameter space to consider; however, dissipation from interactions with the disk powerfully shrinks the relevant phase space.

\subsection{Interactions with the disk} \label{intwdisk}
\cite{Goldreich14} derive an analytic theory for the parametrized disk-damping model used in this paper, assuming the damping timescales are long compared to the resonant timescales.  This is only strictly satisfied (at our lowest masses) in the simulations with $\tau_e = 10^6$ yrs, but their analytical formulae nevertheless provide a useful lens through which to interpret our results.

Eccentricity damping acts to damp the system, not necessarily to $e=0$, but rather to the nearest fixed point in phase space.  If planets are far from resonance, so $e_f \approx 0$ (Eq.\:\ref{ef}), then this indeed damps systems onto circular orbits.  On the other hand, near a resonance, eccentricity damping will move the system to the forced eccentricity, i.e., the most stable configuration.  

At the same time, disks that damp eccentricities should also induce radial migration.  We implement a simple migration scheme by adding a force in our N-body simulations counter to the particle's velocity that continually saps orbital energy \citep{Papa00},
\begin{equation}
\vec{{\bf F_a}} = - \frac{\vec{{\bf v}}}{2 \tau_a},
\end{equation}
where $\vec{{\bf F_a}}$ is the semimajor-axis-damping force vector acting on a planet, and $\vec{{\bf v}}$ is the body's velocity vector.  This yields an orbit-averaged exponential decay of the semimajor axis on a timescale $\tau_a$.  Following \cite{Lee02}, we impose a simple proportionality relation $\tau_a = K \tau_e$, with a nominal value of $K=100$.  This is consistent with the theory for planets embedded in a disk \citep{Arty93, Ward97} (Type I migration), and may apply to planets that have opened a partial gap.  For simplicity, our simulations apply the migration force only to the outer planet, so $\tau_a$ shoud be viewed as the timescale for {\it relative} migration, which is the relevant quantity for the resonant evolution \citep[e.g.,][]{Goldreich14}.  In order for differential migration to lead to capture in resonance, planets must start wide of resonance ($\Delta > 0$) and migrate convergently, rather than vice-versa \citep[e.g.,][Sec. 8.12 and 8.13]{Murray99}.  In such a case, convergent migration pushes planets deeper into resonance on a timescale $\tau_a$, i.e., it moves the fixed point to the right in Fig.\:\ref{resfig} (see Eq.\:\ref{ef}).  

This would likely eventually lead to instability, but eccentricity damping provides a balancing effect, by pushing the stable conjunction longitude away from $\phi = 0$.  This is in some ways analogous to how dissipation in the Earth, coupled with its rotation, causes the tidal bulge raised by the Moon to lead the axes connecting the two bodies.  In both cases, this asymmetry allows the objects to secularly exchange angular momentum, and causes the two bodies to separate from one another.  This ``resonant repulsion" has been quantitatively investigated by \cite{Greenberg81} and more recently by \cite{Lithwick12} and \cite{Batygin13} to explain the overabundance of planet pairs wide of resonance in the Kepler mission's sample.  This orbital divergence causes $\Delta$ to increase, which would move the fixed point toward the origin in Fig.\:\ref{resfig}, following Eq.\:\ref{ef}.  Because resonant repulsion is stronger closer to resonance, the system approaches an equilibrium forced eccentricity $e_\text{eq}$ where the migration is balanced by resonant repulsion.  This occurs at 
\begin{equation} \label{K}
e_\text{eq} \sim \Bigg(\frac{\tau_e}{\tau_a}\Bigg)^{1/2} = K^{-1/2},
\end{equation}
where in general, this equlibrium eccentricity will depend not only on the assumed K, but also on the mass ratio between planets, and the balance of energy and angular momentum dissipation employed for the eccentricity damping \citep[see ][]{Papa05, Goldreich14}.  Regardless of the exact values, dissipation shrinks an otherwise large phase space of initial conditions to tractable equilibrium configurations amenable to numerical investigation.

In such systems acted on by migration and eccentricity damping, \cite{Goldreich14} further find that as one decreases the perturbing planet's mass beyond a critical threshold, the fixed point changes stability and pushes the system away from the forced eccentricity.  In their approximate analytical model, this can cause the system to lose the resonance, in which case the planets migrate through the resonance configuration and do not capture into the equilibrium configuration discussed above.  We do not see such outcomes in our simulations.  This may be due to our working outside their regime of approximation, i.e., we integrate equal-mass planets, and in some cases the damping timescales are comparable to the resonant (libration) timescales.  

\subsection{Stochastic Migration} \label{stochastic}
The discussion above assumes a smooth proto-planetary disk which leads to smooth migration and eccentricity damping forces.
Real disks are expected to be turbulent \citep{Nelson05} either due to the magneto rotational instability (MRI) or the gravitational instability, and this can cause planets to migrate \citep[e.g.,][]{Nelson05,Johnson06, Rein09}.  

The precise level of turbulence is still subject to debate and might depend on a variety of factors such as surface density, surface density gradient, cooling rate, local ionization fraction and dust properties, but an exhaustive discussion of these effects go beyond the scope of this paper.
Here, we parametrize the effect of a non-uniform surface density due to any kind of turbulence with the magnitude of density fluctuations in the disk only:
\begin{equation}
\delta_\Sigma \equiv \Delta \Sigma / \Sigma.
\end{equation}
In numerical simulations, this value has been evaluated to be $\approx 0.1$ in the case of MRI turbulence, in the case of dead zones it is weaker, perhaps of order $0.01$, with significant uncertainties remaining \citep{Nelson05, Oishi07, Nelson10, Gressel11, Yang12}.  A value closer to unity can be expected from strongly gravitationally unstable disks \citep{Rice11}.  

The specific force exerted by a density perturbation\footnote{Note that this expression is independent of the perturbation size $a_\Sigma$ since the mass of the perturbation scales as $M_{\delta\Sigma} \sim \delta_\Sigma \Sigma a_\Sigma^2$ and the force on a planet scales as  ${G M_{\delta\Sigma}}/{a_\Sigma^2}$.} on a nearby planet can then be written as
\begin{equation}
F_{\delta\Sigma} \approx \frac12 \pi G \delta_\Sigma \Sigma
\end{equation}
For many types of instabilities in the disk, absent long-lived coherent structures such as vortices, one might expect the correlation time of density perturbations to be comparable to the orbital timescale, $\tau_c \approx n^{-1}$.
Having the magnitude of random forces and their correlation time, we can define a diffusion coefficient $D$ to completely describe the effects of random forces:
\begin{equation}
D = 2 F_{\delta\Sigma}^2 \tau_c 
= \frac12 \pi^2 G^2 \delta_\Sigma^2 \Sigma^2 n^{-1}.
\end{equation} 
Using the results of \cite{Rein09}, we can now estimate the distance $\Delta a$ that a planet migrates in a time $t$ purely due to stochastic forces:
\begin{eqnarray}
\Delta a = \sqrt{ 4 {Dt}/{n^2} }
\end{eqnarray}
Note that the $\sqrt{t}$ dependence implies that we are modeling this process as a simple random walk in $a$. 
Assuming a power-law dependence with distance, $\Sigma \propto r^{-\gamma}$ ($\gamma < 2$), with a sharp cutoff at $r = R_c$, and relating its normalization to the mass of the disk (see Eq.\:\ref{diskmass}), we can write down the relative change in $a$ in a compact form as a function of time, disk mass and magnitude of density fluctuations:
\begin{eqnarray} \label{stoch}
\frac{\Delta a}{a} \sim \sqrt{\pi} (2-\gamma) \delta_\Sigma \left(\frac{M_d}{M_\star}\right) \left(\frac{a}{R_c}\right)^{2-\gamma} \left( \frac{t}{t_\text{orb}}\right)^{1/2}.
\end{eqnarray}
Planets close in are affected more than planets further out due to their shorter dynamical timescale.  Assuming the value for $\gamma$ traditionally assumed for the minimum-mass solar nebula of 1.5 \citep{Hayashi81}, and using a conservative value of $\delta_\Sigma=0.01$, an HL Tau disk mass of $M_d/M_* \approx 0.25$ \citep{Kwon11} and a period for the innermost planet of $\sim 100$ yrs, we find that within one million yrs, the planet can move by ${\Delta a}/{a} \approx 0.1$.  

This result suggests that stochastic forces from a turbulent disk as massive as that of HL Tau could move around planets significantly during their lifetime.  However, stochastic migration could be mitigated if planets open deep gaps in the gas disk \citep{Rein09} and, depending on the putative planets' masses, other migration mechanisms (e.g., Type I \citealt{Ward97}) could dominate the migration.  In any case, with ALMA and other observatories putting limits on several of the factors in Eq.\:\ref{stoch}, it will be an exciting task to constrain the effects of turbulent migration directly from disk observations \citep[see, e.g.,][]{Hughes11, Simon11, Shi14, Simon15}.

\subsection{Initial conditions} \label{ic}
If planets migrate smoothly toward a particular resonance, the system should remain near the fixed point as the forced eccentricity rises from $\approx 0$ far from the resonance to the equilibrium eccentricity.  In this scenario, when planets are near resonance, initially circular orbits are therefore not the appropriate initial conditions.  They unphysically result in large-amplitude oscillations in both eccentricity and conjunction direction $\phi$ (Fig.\:\ref{resfig}).  This explains why resonances lowered our mass limits in Figs.\:\ref{5planet} and \ref{4planet}, which used initially circular orbits.

Conversely, initial conditions at the fixed point should promote stability and raise our mass limits.  While we have so far considered a simplified dynamical model to gain intuition, in the real problem, each orbit will have its own forced eccentricity that will depend on the different planet masses, damping timescales and proximities to various resonances.  It is therefore impractical to appropriately assign initial conditions.  Instead, we allow the eccentricity damping to move the system to the fixed point by numerically migrating the planets into equilibrium configurations ahead of the ``start" of our simulations.

Of course, if the planetary masses are large enough that the separations correspond to less than the 3.5 $R_H$ limit (see Fig.\:\ref{5planet} and Eq.\:\ref{2planet}), the instability time is short, so before planets can migrate into the stable resonant configuration, the system will disperse.  However, the planets are growing throughout this process.  It is therefore plausible for planets to migrate into resonance at lower masses (when their separations correspond to more than 3.5 $R_H$) and grow together in this stable configuration beyond the limit imposed by Eq.\:\ref{2planet}.  This should provide a robust upper limit to the masses. 

\begin{figure}[!ht]
\includegraphics[width=\columnwidth]{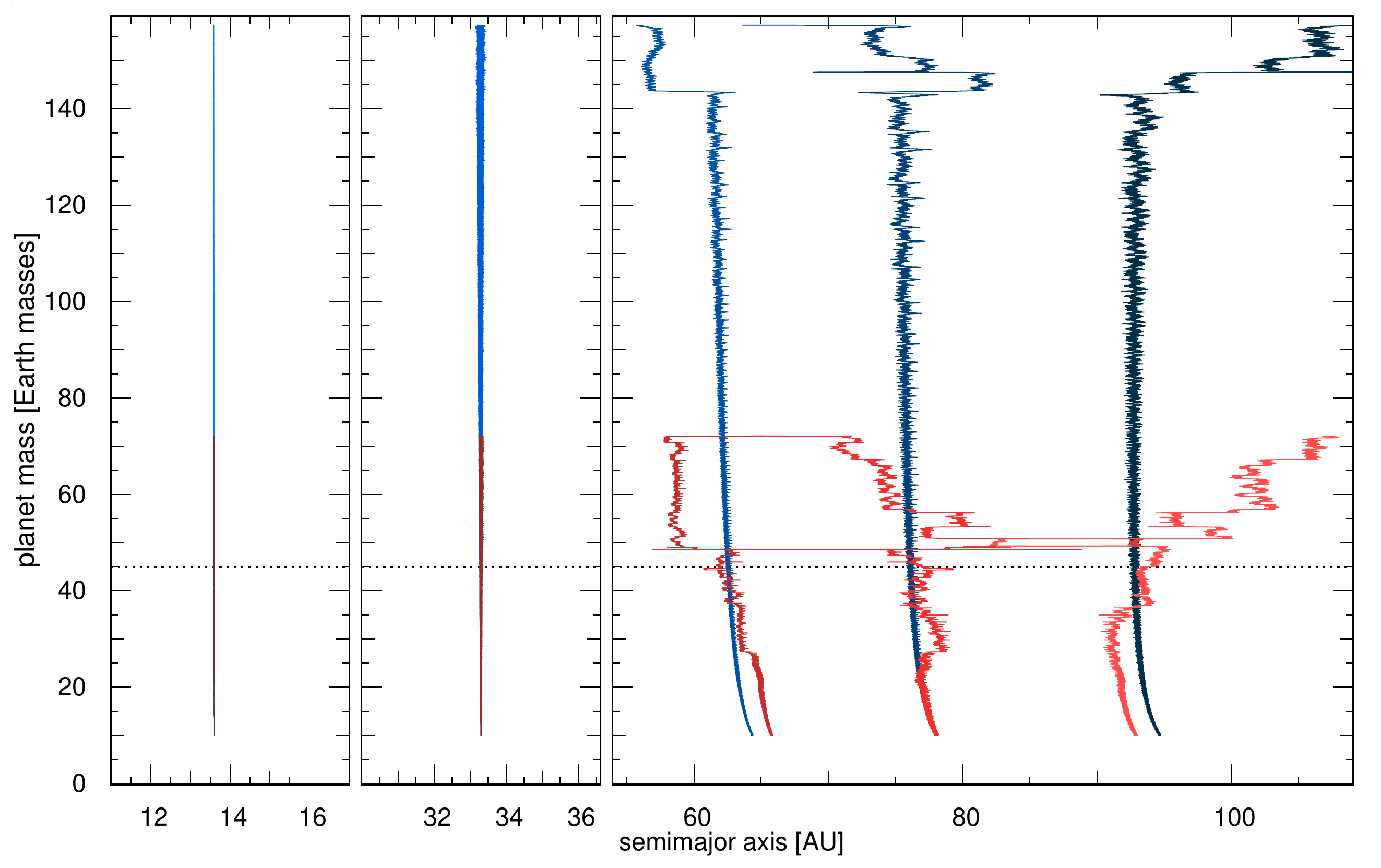}
\caption{\label{growres} Two systems of planets, growing from $10 M_\oplus$ cores (time increases upward).  The outer three blue planets are in a chain of 4:3 resonances, while the red planets are not.  The horizontal dotted line represents the mass threshold above which the orbital separations represent less than 3.5 $R_H$, and non-resonant systems (red) should rapidly destabilitize.  The resonant (blue) system can grow to larger masses through the resonance mitigating the effects of close encounters.}
\end{figure}

This effect is illustrated in Fig.\:\ref{growres}, which shows a simulation of cores growing from $10 M_\oplus$ (time increases upward in the plot).  The system of planets in blue has been migrated into a chain of 4:3 resonances among the outer three bodies, while the system of red planets has not.  Once the planets grow above the dotted horizontal line of 45 $M_\oplus$ (the $3.5 R_H$ boundary for these separations), the non-resonant planets (red) destabilize on the short timescales on which conjunctions occur (Sec.\:\ref{2planet}), while the blue planets continue to grow, protected from close encounters by the resonance.  Eventually the masses grow to the point that kicks at conjunctions knock the planets out of resonance, and swift instability ensues.

We therefore begin with giant planet cores of $10 M_\oplus$, and migrate them into the closest resonant configuration.\footnote{In the $\tau_e = 10^4$ case, the migration time is fast enough that we have to start with $M = 40M_\oplus$ in order to capture into resonance.  See \cite{Ogihara13} for how the mass required for capture is related to the migration timescale.}  In our nominal 5-planet case, both of the outermost pair of planets are near resonance, with period ratios of 1.29 and 1.32, respectively.  We therefore start them at wider separations\footnote{We numerically selected our semimajor axes to yield the observed orbital radii at the end of this ``warmup" of initial conditions (to within 2.5 AU, or half the image's spatial resolution).}, and apply inward migration on the outermost planet at a rate $\tau_a = 100 \tau_e$ until they capture in a chain of 4:3 resonances.  Once the respective resonant angles librate with small amplitude, we adiabatically (on a timescale 10 times longer than the libration period) raise the masses to the value in question.  From this point, we then measure the time to instability as before.

One should ask whether this warm-up time is significant.  We are effectively assuming that planets start in the resonance fully formed, but planets take time to grow, and the equilibrium configuration takes time to set up.  How much time this represents depends on uncertain paramaters such as how close the planets start initially, how the migration timescales vary as the planet grows in mass, etc.  A coherent picture for a system like HL Tau would have to consider these issues.  For simplicity, we sidestep this complication, effectively assuming that the time for planets to grow and for orbits to migrate into resonance is much shorter than the age of the system.  This means that our estimates provide a conservative {\it lower} limit for how large planets can grow while in resonance.

\subsection{Maximum masses in resonant configurations} \label{maxres}
\begin{figure}[!ht]
\includegraphics[width=\columnwidth]{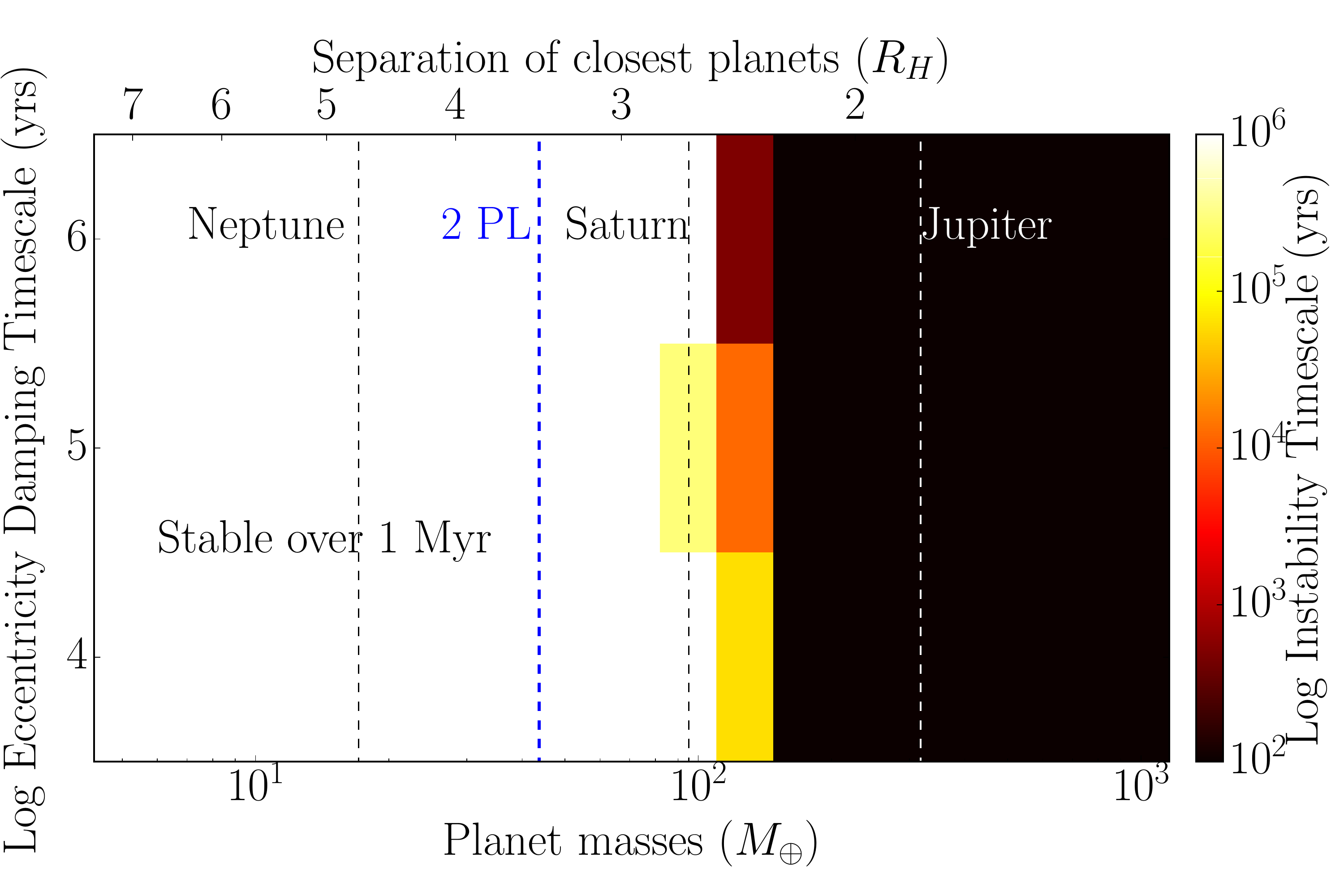}
\caption{\label{5pres} Maximum masses for planets in the HL Tau system.  This assumes a scenario where planets grow while in equilibrium resonance configurations (see Sec.\:\ref{ic} for details).}
\end{figure}

The results for our nominal five-planet case are shown in Fig.\:\ref{5pres}.  Areas in white captured into resonance and were stable for 1 Myr, while areas in black were too unstable to even reach an equilibrium configuration.  The transition is abrupt, and there is only a slight dependence on $\tau_e$, which suggests that we can robustly set a mass threshold of $M \approx 95 M_\oplus$, or about one Saturn-mass.  We reiterate that this represents the maximum mass that can survive for at least 1 Myr in a resonant configuration.  In reality, planets take time to grow and migrate into resonance, so this implies the planet masses could be somewhat greater than our estimate (how much greater depends on uncertain parameters).  This can be seen in Fig.\:\ref{growres}, where the planets grow $\approx 40\%$ beyond a Saturn mass before destabilizing.  We also point out that the resonant configuration pushed the mass limit significantly beyond the 3.5 $R_H$ criterion (dashed blue line in Fig.\:\ref{5pres}), see Eq.\ref{2planet}).  

Performing the same exercise on our more widely separated 4-planet case, we find that planets can grow together in resonance to at least 230 $M_\oplus$.

An important consequence of resonant interactions is that they force eccentricities that will not be damped away.  If gaps in nascent systems are observed to be eccentric, and the period ratios are consistent with a resonant configuration (Sec.\:\ref{errors}), that would provide tantalizing evidence for the interpretation that planets are carving the gaps, and would break the degeneracy as to whether planets are resonant or not.  

The magnitude of the equilibrium eccentricities in our assumed model will depend on both $K$ (Sec.\:\ref{intwdisk}) and the mass ratios between planets \citep{Papa05, Goldreich14}.  In our simulations with $K=100$ and equal masses, we find the equilibrium eccentricities for the gap 3 and gap 5 planets are approximately $0.015$, and $e \approx 0.03$ for the gap 4 planet, independent of $\tau_e$ and the assumed mass.  A survey of the parameter space is beyond the scope of this study, but observed eccentric gaps would clearly motivate such a study.  Smaller values of the uncertain parameter $K$ would yield larger equilibrium eccentricities (see Eq.\:\ref{K}).

The above discussion implicitly assumes that a planet's orbital eccentricity is communicated to the gap it carves.  One might expect that, as long as the gap carving time is short compared to the timescale on which the planet's orbit changes, this should approximately hold true.  This is certainly an important question meriting future study.

An additional complicating factor in the case of HL Tau is that the massive disk should cause the pericenters to precess on timescales ($\sim$ a few orbits, see Sec.\:\ref{prec}) that are likely comparable or faster than the timescale on which planets carve gaps ($\sim 10-100$ orbits \citealt{Fouchet10}).  This would smear out any eccentric signature into a circular gap.  This provides a strong motivation for probing more typical, low-mass disks, $M_D/M_\star \lesssim 1\%$, where the precession periods are likely longer than the timescale for gap opening, and eccentric features could be preserved.

\section{Conclusions}
We divide our conclusions in two.  First we summarize how one might analyze the dynamical stability of an observed system, and then we recapitulate our findings for HL Tau, assuming the gaps are carved by planets.
\subsection{Dynamically constraining ALMA systems} \label{proc}
In this paper, we have provided a framework for understanding the stability of general planetary systems interacting with a gas disk.  The short disk lifetimes ($\lesssim 3$ Myr) over which one might detect nascent systems allow for particularly simple estimates.\footnote{The maximum masses so obtained also apply over longer timescales, they just become less informative.  On Gyr timescales, secular interactions \citep[e.g.,][]{Lithwick11, Wu11, Batygin15} and slower chaotic diffusion can set much more stringent mass limits, at the cost of considerable complexity.}  We now outline a simplified workflow for analyzing a new system, and reference the more detailed discussion in the body of the paper.  These dynamical constraints can easily be incorporated to guide the design and interpretations of both observers and those simulating gap opening in observed systems.

Our treatment has assumed equal-mass planets.  In reality, our procedure effectively estimates the maximum masses $M_\text{crit}$ for the most closely spaced planets in the system (in terms of $\Delta a/a$).  In principle, a similar analysis carried out for the progressively more separated pairs of planets in the system should yield reliable results.  It would be an interesting extension of this work to test how accurately these simple criteria can be applied in this iterative fashion.  We now summarize a simple procedure for analyzing a system's stability.

\begin{itemize}
	\item The base estimate for $M_\text{crit}$ is given by inserting the minimum fractional separation among planet pairs $(\Delta a/a)_\text{min}$ in Eq.\:\ref{2planet}.  If these planets are not near resonances (see below), this provides a stringent upper mass limit for the planets in question, which is independent of the assumed interactions with the disk.  
\end{itemize}
One can then proceed to refine this limit, which only rigorously applies in the case of a two-planet system.  
\begin{itemize}
	\item If planets on either side of the closest two have $\Delta a /a \lesssim 4 (\Delta a/a)_\text{min}$, and the eccentricity damping timescale is long compared to the time betwen conjunctions (see footnote in Sec.\:\ref{2planet}), planet masses below the above-calculated $M_\text{crit}$ will destabilize over the system's lifetime (Sec.\:\ref{3pl}).  
	\item In this large $\tau_e$ regime, and in the limiting case that adjacent pairs share $\Delta a /a  = (\Delta a/a)_\text{min}$, $M_\text{crit}$ is pushed down to the value given by Eq.\:\ref{3planet}.  In intermediate cases, Eqs.\:\ref{2planet} and \ref{3planet} bracket $M_\text{crit}$ \citep{Chambers96, Chatterjee08, Marzari14}.  
\end{itemize}

Near-resonant orbits will also modify $M_\text{crit}$.  
\begin{itemize}
	\item The locations of resonances are uncertain to within $\sim M_\text{disk}/M_\star$ (Sec.\:\ref{prec}), and the orbital radii to within some observational uncertainty (Sec.\:\ref{errors}).
	\item If, within these uncertainties, the orbits' fractional distances from first-order MMRs are always large compared to $M_\text{crit}/M_\star$ (Sec.\:\ref{resimp}), then $M_\text{crit}$ is a robust estimate.
	\item If first-order MMRs can not be ruled out, one can numerically set an upper mass limit by assuming the planets have reached an equilibrium resonant configuration (Sec.\:\ref{ic})
\end{itemize}

\subsection{HL Tau} \label{hltauconc}
Our main conclusions are summarized in the bullet points below,\footnote{Note added during revision:  An analysis of the HL Tau data has recently appeared \citep{Brogan15}.  They find that gaps 3 and 4 are closer than we assume; using their values would somewhat lower our quoted maximum masses in the 5-planet case.  \cite{Brogan15} also use the disk kinematics to derive a mass that is approximately twice as large ($\sim 1.3 M_\odot$) as inferred by previous studies.  If indeed the star is this massive, and the disk is somewhat less massive than suggested by the modeling of \cite{Kwon11}, this would correspondingly reduce the shifts of resonances from their nominal locations (Sec.\:\ref{prec}), and mitigate the rapid erasure of eccentric gaps by precession (Sec.\:\ref{maxres}).  Remarkably, \cite{Brogan15} find each of the gaps to be offset from center.  It will be an exciting task to determine whether these offsets are due to eccentric gaps and to exploit this information to further probe the planetary hypothesis.}
\begin{itemize}
	\item A disk's mass should shift the resonance locations from integer period ratios by $\sim M_D/M_\star$ $(\sim 10\%$ for HL Tau), but the precise value is uncertain due to theoretical difficulties in evaluating the contribution to orbital precession rates from local material disturbed by the planet (Sec.\:\ref{prec}).  
	\item This theoretical uncertainty, coupled with observational uncertainties in the orbital radii make it difficult to constrain whether planets are resonantly interacting.  Conversely, in closely spaced systems like HL Tau, one cannot rule out resonant configurations (Sec.\:\ref{errors}).
	\item Despite the possible window into the era of planet formation afforded by the HL Tau data, these uncertainties in the dynamical state are an important consideration when using the HL Tau system (or others discovered in the future) as initial conditions for simulations of, e.g., planet scattering.
	\item Observationally detecting eccentric gaps that may be resonant would provide tantalizing evidence that resonantly interacting planets are causing them.  This would additionally break the degeneracy in the system's dynamical state, constrain the pericenter precession rates and provide proxies for initial conditions at the end of planet formation (Sec.\:\ref{maxres}).
	\item The rapid pericenter precession rates induced by massive disks will erase eccentric signatures and generate circular gaps (Sec.\:\ref{maxres}).  This strongly motivates pushing toward less massive, more typical disks with $M_D/M_\star \lesssim 10^{-2}$.
	\item The timescale for stochastic diffusion onto crossing orbits, as well as observational limits on the level of density fluctuations in the disk, can set limits on the level of turbulence in the outer disk regions of systems probed by ALMA (Sec.\:\ref{stochastic}).
	\item Planets can grow to much larger masses than one would naively expect if they capture in resonance at low masses and grow together (Fig.\:\ref{growres}).  This can continue until the masses become large enough that kicks at conjunctions knock the planets out of resonance, leading to swift instability.  It would be valuable to investigate how the threshold mass varies between different first-order resonances, as well as with the assumed dissipation timescales.  If distant planets tend to capture into resonances, this could represent a universal pathway to instability (Sec.\:\ref{ic}).  
\end{itemize}
As an illustration of the framework we have elaborated, we focused primarily on a nominal model with five planets orbiting in the most prominent gaps (Fig.\:\ref{hltauimg}), as well as a 4-planet model where gaps 3 and 4 are generated by a single planet lying between the gaps, with co-orbital material along its path.  In both cases, the inner two planets are largely dynamically decoupled from the system, so the below mass limits apply to the outer planets.  
\begin{itemize}
	\item 5-planet Case (Fig.\:\ref{hltauimg}):  If planets are not resonantly interacting, the maximum masses that allow for stability over the system's age are $\approx 2$ Neptune masses.  If planets grow together while in resonance, masses can reach at least $1$ Saturn mass.  
	\item 4-planet Case:  In the non-resonant case, planet masses can be as large as $\approx 1$ Saturn mass, while growing together in resonance, masses can reach at least 230 $M_\oplus$.
	\item The inner planets are largely decoupled from the system and are consistent (dynamically) with at least several Jupiter masses (Eq.\:\ref{3planet}).  At these large masses, higher-order corrections in the planet-to-star mass ratio become important, and our equations may not be precise.
\end{itemize}
Additional observations will likely produce high resolution maps of the gas disk. Whether gas is observed at the locations of gaps in the dust will additionally constrain the maximum masses of any planets carving them. In addition, comparing the dust and gas distributions can help pinpoint the stage of planetary formation the system is in: during the build-up of giant-planet cores, during the runaway accretion of gas, or in between both phases. Given constraints on the system's age, it may soon be possible to empirically test theoretical expectations \citep[e.g.,][]{Pollack96, Alibert05}.  Putting new observations and simulations in the dynamical context we have presented should prove an exciting and fruitful task.

\bigskip
We would like to thank Yanqin Wu, Norman Murray, Hilke Schlichting, Pawel Artymowicz, Ruth Murray-Clay and Katherine Krettke for insightful discussions.  We are also grateful to the anonymous referee who helped improve this manuscript.  This research was made possible by the Sunnyvale cluster at the Canadian Institute for Theoretical Astrophysics.

\appendix
\section{Appendix A:  Determination of orbital radii from the ALMA image} \label{Amaury}
\cite{Kwon11} used CARMA\footnote{Combined Array for Research in Millimeter-wave Astronomy} observations in a band centered at the same wavelength (1.3 mm) as the ALMA image to estimate the disk's diameter at 235 AU. Additionally, they constrained the PA angle at $136^\circ$ and the inclination of the disc at $40^\circ$. Using commercial software, we drew ellipses onto the press-release image. We adopted a PA angle of $136^\circ$; however, through trial and error, we found that if one assumes the orbits are circular, the aspect ratio between the minor and major axes implies an inclination of the system of $\approx 48^\circ$. This is likely caused by a non circular beam on the ALMA image, which we could not compensate for. The minor axis only served the purpose of centering the ellipses and does not impact the value that interest us: the orbital diameter of each of the gaps, which we estimated using the major axis.

We started by forcing all ellipses to share a common center; the aspect ratio was also kept constant. The width of each ellipse's circumference was set to correspond to the average width of the gaps. This meant that the inner and outer edges both contribute to identifying the middle of the gap. We visually adjusted only the major axes, each independently. A first ellipse was adjusted to one of the gaps (randomly chosen). The order with which subsequent gaps were located, was random. The full procedure was repeated five times. This means that each gap has five orbital diameter estimates. We gathered the set and computed the mean of each gap's diameter and provide an error using the rms of the values. 

In Fig.\:\ref{hltauimg}, gap 5 is visibly not well centered. Because of this, a second series of five adjustments was made, reproducing the previous methodology except that now the center of each ellipse was independently positioned. We obtained consistent results within the error bars, so we conclude that any eccentricity in the outer ring is not significant at the precision level of our approach.

Both sets being in agreement, each gap has ten estimates, which we combined to produce a single set of orbital distances. We thus obtain diameters in arbitrary units, which we denote pt.  To convert to an absolute scale using the CARMA diameter of 235 AU \citep{Kwon11}, we estimated the diameter of the disk by drawing an additional, thin, ellipse, ten times independently. This yielded a diameter of the disk of $1385 \pm 11$pt, which we used to convert the orbital radii into AU in the second column of table~\ref{gapstable}. 

We do not include in our estimates an error incurred by this unit conversion (the absolute scale is uncertain to at least the level of the error in the distance to HL Tau, approximately 15\% \citealt{Torres07}), partly because this uncertainty in absolute scale should not have a strong impact on our results.  The major effect is to change the orbital periods for a given assumed mass for the central star.  But because instability timescales are such a steep function of the {\it relative} separations (Sec.\:\ref{3planet}), a slight change in the number of orbits executed does not qualitatively influence the results.  The important quantities in our analysis are the relative separations, which are well constrained by our method.

\begin{table}[!ht] \label{gapstable}
\caption{Gap locations\footnote{Errors do not include uncertainties in the image's overall scale.  Such a rescaling would have little effect on our results (see Appendix \ref{Amaury}).} as estimated from Figure~\ref{hltauimg}.  }\label{gaps}
\begin{tabular}{lll}
\hline
\hline
gap   & diameters & physical \\
 number  & (arbitrary unit) & orbital radii (AU)\\
  \hline
1 &$160.1\pm2.6$ & $13.6\pm0.2$  \\
2 &$392.8\pm2.2$ & $33.3\pm0.2$  \\
3 &$767.5\pm7.5$ & $65.1\pm0.6$  \\
4 &$911.3\pm4.4$ & $77.3\pm0.4$  \\
5 &$1096\pm11$ & $93.0\pm0.9$  \\
\end{tabular}
\end{table}

\section{Appendix B: Precession induced by a thin axisymmetric disk} \label{precapp}
We consider a razor-thin, axisymmetric disk with a power-law surface density $\Sigma(r)$,
\begin{equation} \label{sigma}
\Sigma(r) = \Sigma_0 \Bigg(\frac{R_c}{r}\Bigg)^\gamma,
\end{equation}
where $r$ is the radial distance, $\Sigma_0$ is a normalization, $R_c$ is a characteristic radius, and $\gamma$ is the power-law index.  If the disk is infinite, one obtains the simple result for the radial force $F_D(r)$ \citep{Lemos91, Evans97},
\begin{equation} \label{infdisk}
F_D(r) = -F_0\Bigg(\frac{R_c}{r}\Bigg)^\gamma,
\end{equation}
where $F_0$ is another normalization.  Such a disk has infinite mass, so we generalize the above solution to a case where one imposes a sharp cutoff at $R_c$.  Specializing to $\gamma < 2$ (one could also add an inner cutoff to avoid divergence at $r=0$ for $\gamma > 2$), the mass of the disk $M_D$ is 
\begin{equation} \label{diskmass}
M_D = \int_{0}^{R_c} \Sigma_0 \Bigg(\frac{R_c}{r}\Bigg)^\gamma (2\pi r) dr = \frac{2\pi \Sigma_0 R_c^2}{2-\gamma}.
\end{equation}
In this case, as in the infinite case, one can elegantly solve the problem by constructing the surface density from thin homoeoids\footnote{The generalization of spherical shells for oblate spheroids.  It can be shown (see, e.g., \citealt{Binney08}) that a mass lying interior to a homoeoid feels no force (a generalization of Newton's famous result).} that are completely flattenened into a 2-D disk.  The result is \citep[see Eq.\:2.157 in ][]{Binney08},
\begin{equation}
\frac{F_D(r)}{m} =\frac{4G}{r} \int_{0}^{r} da \frac{a}{\sqrt{r^2 - a^2}} \frac{d}{da} \int_{a}^{\infty} dr' \frac{r' \Sigma(r')}{\sqrt{r'^2 - a^2}},
\end{equation}
where $G$ is the gravitational constant and $a$ represents a homoeoid's semimajor axis.  If the surface density cuts off at $r = R_c$, $R_c$ replaces the upper limit in the second integral, yielding the correction to Eq.\:\ref{infdisk}
\begin{equation} \label{fgen}
F_D(r) = -F_0 \Bigg(\frac{R_c}{r}\Bigg)^\gamma [1 + \eta(r,\gamma)],
\end{equation}
where 
\begin{equation}
F_0 = \xi \frac{G M_D}{R_c^2},
\end{equation}
\begin{equation}
\xi = (2-\gamma)\frac{\Gamma{[\frac{2-\gamma}{2}]}\Gamma{[\frac{1+\gamma}{2}}]}{\Gamma{[\frac{3-\gamma}{2}]}\Gamma{[\frac{\gamma}{2}}]},
\end{equation}
$\Gamma$ is the Gamma function, 
\begin{equation}
\eta(r, \gamma) = \frac{2-\gamma}{\pi \xi} \Bigg(\frac{R_c}{r}\Bigg)^{1-\gamma}\Bigg[2K\Bigg(\frac{r^2}{R_c^2}\Bigg) - \pi \Gamma \Bigg(\frac{1 + \gamma}{2}\Bigg) H(\gamma, \frac{r^2}{R_c^2})\Bigg],
\end{equation}
and $H$ is the regularized, generalized hypergeometric function
\begin{equation}
\frac{_3H_2\Bigg(\{1/2, 1/2, \frac{\gamma-1}{2}\},\{1,\frac{1+\gamma}{2}\},\Bigg(\frac{r}{R_c}\Bigg)^2\Bigg)}{\Gamma[\frac{1+\gamma}{2}]}.
\end{equation}
From Eq.\:\ref{fgen}, $\eta(r)$, represents the fractional correction due to the disk truncation, relative to the case of an infinite disk (Eq.\:\ref{infdisk}), and $\eta(r)$ is always greater than zero.  This reflects the fact that the outer material that would have contributed a positive force is now missing.  This correction increases as $r$ approaches $R_c$, but is normally small.  For $\gamma = 1.5$, the traditional value for the minimum-mass solar nebula (MMSN, \citealt{Hayashi81}), and for an HL Tau value of $R_c = 117.5$ AU, $\eta \approx 10\%$ at the outermost gap.  

We can now obtain the pericenter precession rate through
\begin{equation} \label{genperi}
\frac{\dot{\varpi}}{n} = -\Bigg[\frac{F_D}{F_K} + \frac{1}{2} r \frac{dF_D/dr}{F_K}\Bigg],
\end{equation}
where $F_K = GM_\star/r^2$ is the Keplerian force from the star.  This yields 
\begin{equation} \label{precrate}
\frac{\dot{\varpi}}{n} = - \frac{M_D}{M_\star}(1-\frac{\gamma}{2}) \xi \Bigg(\frac{R_c}{r}\Bigg)^{\gamma-2}\Bigg[1 + \eta_2(r, \gamma)\Bigg],
\end{equation}
where $\eta_2(r, \gamma)$ is the fractional correction relative to the case of an infinite disk (in which case one should replace $M_D$ according to Eq.\:\ref{diskmass} and $R_c$ becomes a characteristic, rather than a cutoff radius).  Eq.\:\ref{precrate} gives the same functional dependence found by \cite{Rafikov13}, who specialized to the classical Mestel disk, i.e., $\gamma = 1$.

Rather than provide an inscrutable expression for $\eta_2$, we give a qualitative analysis.  We first note that the two quantities in the square brackets of Eq.\:\ref{genperi} have opposite signs; since both $F_D$ and $F_K$ are negative, the first term is positive, while, because $F_D$ approaches zero at large $r$, $dF_D/dr > 0$ and the second term is negative.  We also point out that the final answer for the precession rate (Eq.\:\ref{precrate}) is negative (recall that we have specialized to $\gamma < 2$).  If we now consider the effect of truncating the disk, as mentioned above, this makes $F_D$ more negative, and the effect increases as one approaches $R_c$ (i.e., $dF_D/dr$ decreases, since $F_D$ rises slower).  Thus, the first term in square brackets in Eq.\:\ref{genperi} becomes more positive, and the second less negative, so truncating the disk always acts to enhance the magnitude of $\dot{\varpi}$.

In this case, the fractional correction $\eta_2$ can be quite large, due to the strong effect on the derivative in Eq.\:\ref{genperi}.  Again for $\gamma = 1.5$ and $R_c = 117.5$ AU, $\eta_2 \approx 90\%$ at the outermost gap.  Thus, while this is an important quantitative correction, one nevertheless obtains qualitatively correct lower limit to the precession rate by taking $\eta_2 = 0$, i.e., by assuming an infinite disk for the force law (Eq.\:\ref{infdisk}), but still using the mass of the disk for $M_D$ in Eq.\:\ref{precrate}.

\end{document}